\documentclass[acmsmall,screen]{acmart}

\usepackage{url}
\usepackage{xspace}
\usepackage{booktabs}
\usepackage{makecell}
\usepackage{hyperref}
\usepackage{graphicx}
\usepackage{colortbl}
\usepackage{multirow}
\usepackage{subfigure}
\usepackage{subcaption}
\usepackage{tablefootnote}
\usepackage{color, xcolor}
\usepackage{amsthm,amsmath,amsfonts}
\usepackage{pifont}
\usepackage{enumitem}
\usepackage{listings}
\usepackage{fontawesome5}
\usepackage{algorithm}
\usepackage{algpseudocode}
\usepackage{algorithmicx}
\usepackage[most]{tcolorbox}
\tcbuselibrary{breakable, skins}
\usepackage{svg}
\usepackage{xspace}
\newcommand{\toolname}{\textsc{SGAgent}\xspace}
\newcommand{\ie}{\textit{i.e.,}\xspace}
\newcommand{\eg}{\textit{e.g.,}\xspace}

\newcommand{\summary}[2]{
\begin{center}
\begin{tcolorbox}[leftrule=0mm,toprule=0mm,bottomrule=0mm,rightrule=0mm,left=1pt,right=2pt,top=0pt,bottom=0pt,breakable]
\textbf{Answer to RQ{#1}:}
{#2}
\end{tcolorbox}
\end{center}
}
\newcommand{\find}[2]{
\begin{tcolorbox}[toprule=0mm,bottomrule=0mm,left=1pt,right=2pt,top=2pt,bottom=2pt,breakable]%
\em #1
\end{tcolorbox}
}

\newtcolorbox[auto counter]{promptbox}[2][]{
  breakable,
  colback=gray!5,
  colframe=gray!60,
  coltitle=black,
  fonttitle=\bfseries,
  enhanced,
  sharp corners,
  boxrule=0.8pt,
  left=6pt,
  right=6pt,
  top=6pt,
  bottom=6pt,fontupper=\small,
  title={Listing~\thetcbcounter: #2},
  #1
}

\AtBeginDocument{
  \providecommand\BibTeX{{
    \normalfont B\kern-0.5em{\scshape i\kern-0.25em b}\kern-0.8em\TeX}}}

\setcopyright{acmcopyright}
\copyrightyear{2024}
\acmYear{2024}
\acmDOI{1}

\acmJournal{TOSEM}
\acmVolume{1}
\acmNumber{1}
\acmArticle{1}
\acmMonth{1}

\begin{document}

\title{\toolname{}: Suggestion-Guided LLM-Based Multi-Agent Framework for Repository-Level Software Repair}

\author{Quanjun Zhang}
\email{quanjunzhang@njust.edu.cn}
\orcid{0000-0002-2495-3805}
\affiliation{
\institution{School of Computer Science and Engineering, Nanjing University of Science and Technology}
\city{Nanjing}\country{China}
}

\author{Chengyu Gao}
\email{502023320004@smail.nju.edu.cn}
\orcid{0000-0002-9930-7111}

\author{Yu Han}
\email{522024320053@smail.nju.edu.cn}
\orcid{0000-0002-9930-7111}

\author{Ye Shang}
\email{yeshang@smail.nju.edu.cn}
\orcid{0009-0000-8699-8075}

\author{Chunrong Fang}
\authornotemark[1]
\email{fangchunrong@nju.edu.cn}
\orcid{0000-0002-9930-7111}

\author{Zhenyu Chen}
\email{zychen@nju.edu.cn}
\orcid{0000-0002-9592-7022}
\affiliation{
\institution{State Key Laboratory for Novel Software Technology, Nanjing University}
\city{Nanjing}\country{China}
}

\author{Liang Xiao}
\authornote{Chunrong Fang and Liang Xiao are the corresponding authors.}
\email{xiaoliang@mail.njust.edu.cn}
\orcid{0000-0003-0178-9384}
\affiliation{
\institution{School of Computer Science and Engineering, Nanjing University of Science and Technology}
\city{Nanjing}\country{China}
}

\begin{abstract}
The rapid advancement of Large Language Models (LLMs) has led to the emergence of intelligent agents capable of autonomously interacting with environments and invoking external tools.
Recently, agent-based software repair approaches have received widespread attention, as repair agents can automatically analyze and localize bugs, generate patches, and achieve state-of-the-art performance on repository-level benchmarks (\eg, SWE-Bench).
However, existing software repair approaches usually adopt a \textit{localize-then-fix paradigm}, jumping directly from ``where the bug is'' to ``how to fix it'', leaving a fundamental reasoning gap.

To this end, we propose \toolname{}, a \underline{S}uggestion-\underline{G}uided multi-\underline{Agent} framework for repository-level software repair, which follows a \textit{localize-suggest-fix} paradigm. 
Specifically, \toolname{} introduces a suggestion to strengthen the transition from localization to repair.
The suggester starts from the buggy locations and incrementally retrieves relevant context until it fully understands the bug, and then provides actionable repair suggestions. 
Moreover, we construct a Knowledge Graph (KG) from the target repository and develop a KG-based toolkit to enhance \toolname{}'s ability to enhance global contextual awareness and repository-level reasoning. 
Based on these components, three specialized sub-agents in \toolname{} (\ie localizer, suggester, and fixer) collaborate to achieve automated end-to-end software repair.
We evaluated \toolname{} on the SWE-Bench-Lite benchmark.
Experimental results show that \toolname{} with Claude-3.5 achieves 51.3\% repair accuracy, 81.2\% file-level, and 52.4\% function-level localization accuracy with an average cost of \$1.48 per instance, outperforming all baselines using the same base model.
Moreover, \toolname{} generalizes well across different base LLMs, further reaching a 60.7\% resolution rate with Claude-4.
When extended to vulnerability repair, \toolname{} achieves a 48.0\% resolution rate on VUL4J and VJBench, demonstrating strong generalization across tasks and programming languages.

\end{abstract}




\maketitle

\section{Introduction}
\label{sec:introduction}

Recently, Large Language Models (LLMs) have achieved remarkable progress across a wide range of software engineering tasks~\cite{zhang2023survey,hou2023large}, including code generation~\cite{jiang2024survey,du2024evaluating}, test generation~\cite{zhang2025large,shang2025large}, and software repair~\cite{zhang2024systematic,zhang2023gamma}.
These advancements demonstrate the growing potential of LLMs to understand, generate, and manipulate source code in increasingly complex scenarios.
However, most existing studies primarily concentrate on simplified, function-level tasks, where the repair or generation scope is limited to a single function or file~\cite{agentless,KGCompass}.
Such settings, while effective for controlled evaluation, fail to capture the repository-level dependencies and contextual complexities inherent in real-world software development.

To bridge this gap, the research community has recently turned its attention to repository-level software repair, which has rapidly emerged as one of the most challenging and promising directions in LLM-based software engineering~\cite{AutoGPT,openhandsArticle,schmidgall2025agent,zhao2024expel}.
In particular, the popular benchmark SWE-Bench~\cite{swe-bench} has been introduced to evaluate LLMs' capabilities for resolving real GitHub issues with full repository access.
Each task in SWE-Bench includes a natural-language issue description and its corresponding codebase, requiring LLMs to perform repository-level code reasoning to solve software-evolution challenges such as bug fixing and feature implementation.
Given the computational overhead of running the full benchmark, the authors further released SWE-Bench Lite, a curated, lightweight subset that has become one of the most widely adopted benchmarks for evaluating LLMs on repository-level software repair tasks.

To address the challenging real-world software engineering problems posed by SWE-bench, researchers have explored two primary paradigms: procedural~\cite{agentless,KGCompass} and agentic~\cite{AutoGPT,openhandsArticle} ones.
In particular, agentic approaches~\cite{AutoGPT,openhandsArticle} equip LLMs with external tools (\eg, file editing, code retrieval, and test execution) for iterative planning and environment interaction, enabling LLMs to autonomously navigate and modify large codebases in a manner akin to human developers.
In contrast, procedural approaches~\cite{agentless,KGCompass} integrate LLMs into predefined repair pipelines, and operate sequentially through multiple stages, such as fault localization, patch generation, and patch validation.
Both paradigms have repeatedly advanced the state of the art on SWE-Bench, demonstrating the substantial potential of LLM-based repository-level software repair.

Despite recent promising results, existing software repair approaches mainly follow a \textit{localize-then-fix} paradigm, \ie first identifying suspicious code elements and directly feeding them to LLMs to generate patches.
However, jumping directly from fault localization to patch generation introduces several fundamental challenges. 
(1) Objective mismatch.
Fault localization answers where the fault might be, whereas patch generation concerns how to modify the codebase.
These two tasks require different semantic granularity and contextual understanding, and thus localization results alone do not provide sufficient guidance for reliable patch synthesis.
In particular, simply providing LLMs with localized code snippets omits critical repository-level semantics, such as cross-file dependencies, meaning that knowing where the bug occurs does not provide enough information to determine what and how to change.
(2) Sensitivity on localization results.
The success of downstream patch generation largely depends on the quality of localization results~\cite{zhang2023gamma}. 
However, when the localization results are inaccurate, directly passing identified snippets to patch generation may propagate and even amplify errors in the downstream repair process.
(3) Missing intermediate planning.
Without an explicit planning phase between localization and repair, LLMs tend to directly apply edits to the nearest suspicious snippet, relying on trial-and-error rather than reasoning about why and how the code should be modified. 
As a result, generated patches often overfit to the identified snippets, leading to redundant exploration and superficial fixes. 
Moreover, such opaque repair behavior reduces trustworthiness and hinders practical adoption in real-world development workflows, where developers require interpretable and justifiable repair suggestions.

\textbf{This Paper}.
To address the above issues, we propose \toolname{}, a suggestion-guided, repository-level software repair approach that mimics human debugging practice based on multi-agent collaboration.
In real-world debugging scenarios, developers typically follow a three-step process: (1) analyze the issue behavior and localize suspicious code snippets; (2) reason about interactions between the suspicious region and other files, and devise a possible fix plan; and (3) attempt to modify the codebase and return a correct patch.
Inspired by this expert practice, we design \toolname{} with three specialized agents: a fault localizer that identifies candidate fault regions; a repair suggester that performs backward analysis to convert coarse localization signals into actionable, repository-aware edit plans; and a patch generator that executes the plan and synthesizes patches. 
Importantly, the suggest stage closes the gap between where and how to fix issues by reinterpreting localization outputs before patching, thereby mitigating the challenge of imperfect localization and enabling interpretable, step-wise repair.

Furthermore, to emulate how developers interact with IDEs to navigate the complex structure of codebases, we design and implement a fine-grained knowledge graph architecture and an API-based toolkit comprising 3 entity types, 7 relationship types, and 14 tools. 
This framework effectively covers nearly all static and dynamic program relationships (\eg, function calls, dependencies). Based on this, we not only leverage the knowledge graph to retrieve repository information during the localization phase, but also equip the suggester with the toolkit, enabling it to trace back related code from the identified buggy locations, provide repair suggestions and refine repair objectives.

We evaluate \toolname{} on the SWE-Bench-Lite dataset.
Experimental results demonstrate that \toolname{} achieves a 51.3\% issue resolution rate, outperforming existing baselines under the same LLM setting.
Moreover, when implemented with Claude-4, \toolname{} further improves the resolution rate to 60.7\%, demonstrating that our suggestion-guided multi-agent framework can effectively generalize across different backbone models.
To provide a more rigorous evaluation, we additionally conduct experiments on SWE-Bench-Verified, where \toolname{} resolves 327 out of 500 instances and achieves a 65.4\% resolution rate, confirming its effectiveness on high-quality, human-validated repository-level repair tasks.  
We also evaluated the vulnerability repair capability of \toolname{} on combined VUL4J and VJBench datasets. 
The results show that \toolname{} achieves an average resolution rate of 48.0\%, demonstrating its strong generalization ability to security-related repair scenarios.

\textbf{Novelty \& Contributions.}
To sum up, the main contributions of this paper are as follows:

\begin{itemize}
    \item 
    We propose \toolname{}, a novel multi-agent framework for automatically repairing real-world repository-level software bugs. 
    Drawing inspiration from real-world software repair processes, we design three agents with distinct roles, including localizer, suggester, and fixer. 
    They collaborate with each other to complete bug localization and repair.
    
    \item 
    We replace the localize-then-fix paradigm with the locate-suggest-fix paradigm that mimics expert debugging expertise. 
    By introducing an extra suggest stage before generating patches, \toolname{} resolves 13.3\% more instances on the SWE-Bench-Lite~\cite{swe-bench} dataset.

    \item 
    We are the first to develop a complex, scenario-oriented toolkit built upon a self-implemented fine-grained knowledge graph. The toolkit comprises 14 specialized tools, each designed to address specific repository-level software repair scenarios, and is tightly integrated into \toolname{}. This integration significantly enhances \toolname{}’s context retrieval and repository understanding capabilities, resulting in a 3.1\% improvement in overall performance.
    
    \item 
    We evaluate \toolname{} on the SWE-Bench dataset and compare its performance against state-of-the-art approaches. Experimental results demonstrate that, under the same underlying LLM setting, \toolname{} outperforms all baselines with a 51.3\% resolve rate, an 81.2\% file-level and 52.4\% function-level localization accuracy, while maintaining a relatively low repair cost of \$1.48 per instance (ranks third among all baselines). 
    This highlights the significant potential of the repair agent and provides valuable guidance for future research.
    
    \item 
    To facilitate reproducibility and further research, we release the full implementation of \toolname{}, including the source code, experiment configurations, and data processing pipeline. The project is openly available in our public repository~\cite{sgagent}.
\end{itemize}

\section{Motivation}

\begin{figure*}[htbp]
    \centering
    \includegraphics[width=1\linewidth]{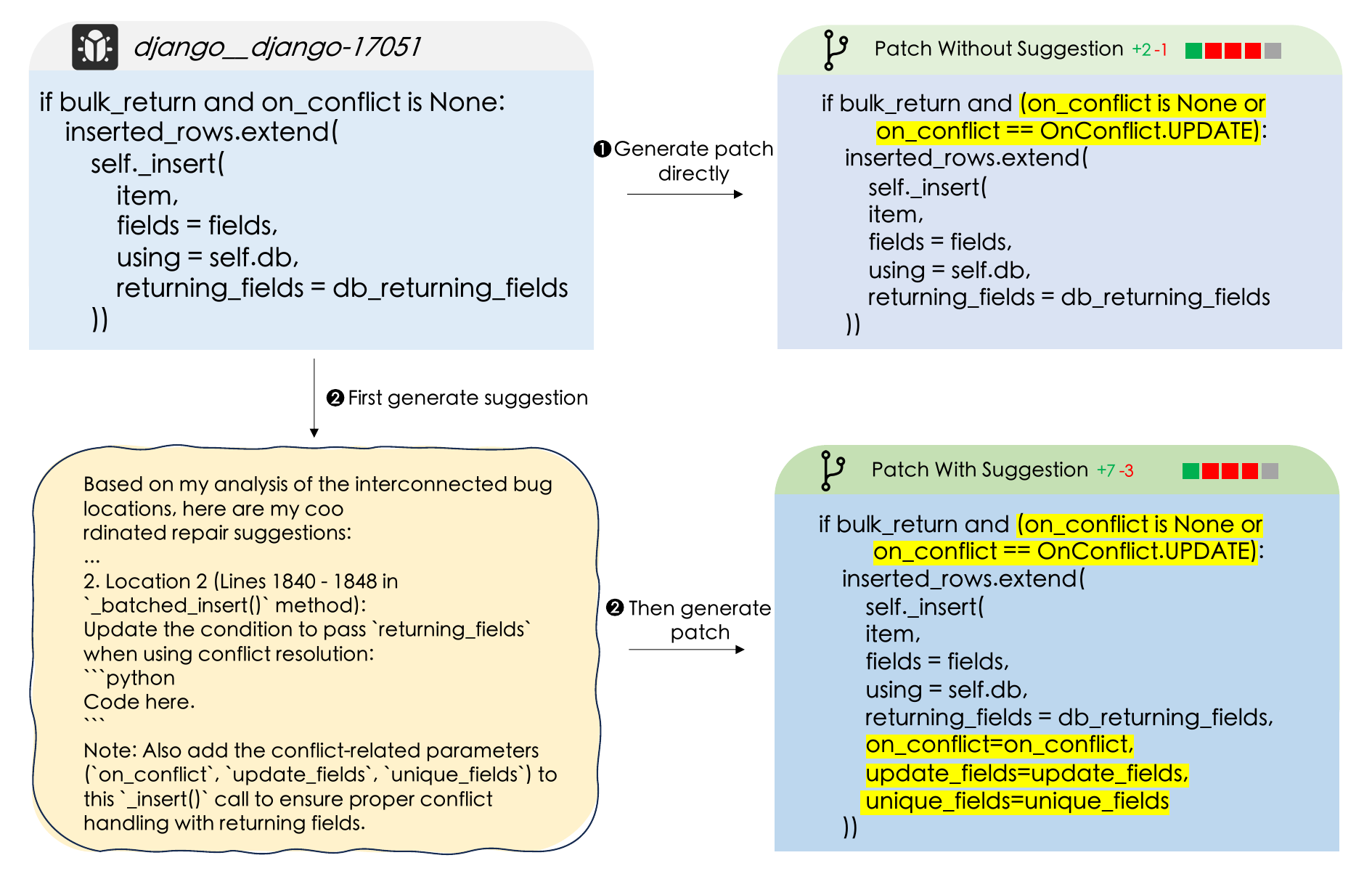}
    \caption{A motivation example of \toolname{}.}
    \Description{A figure showing a motivation example for the tool.}
    \label{fig:motiv}
\end{figure*}

To illustrate the motivation behind \toolname{}, we present a motivation example from the SWE-Bench benchmark.  
As shown in Figure~\ref{fig:motiv}, the bug instance \textit{django\_\_django-17051} lies in the failure to properly pass conflict-handling parameters \textit{on\_conflict, update\_fields, unique\_fields} from \textit{\_batched\_insert()} to \textit{\_insert()}, causing the underlying logic to incorrectly handle \texttt{ON CONFLICT DO UPDATE} semantics despite modified conditions.

We experimented with two repair strategies: the first directly generates a patch after localization, while the second first uses a model to generate repair suggestions based on the localization results, and then the repair model synthesizes both the localization output and the suggestions to produce the final patch.
The results show that without repair suggestions, \textit{django\_\_django-17051} cannot be correctly fixed. The generated patch simply modifies the conditional statement \textit{if bulk\_return and on\_conflict is None} to \textit{if bulk\_return and (on\_conflict is None or on\_conflict == OnConflict.UPDATE)}. Although this change superficially aligns with the intention of allowing UPDATE conflicts and return IDs, it fails to handle the deeper semantic dependencies between related functions. Specifically, the \textit{\_insert()} call neglects to pass critical conflict-related parameters \textit{on\_conflict}, \textit{update\_fields}, and \textit{unique\_fields}, which are necessary for the underlying insertion logic to correctly execute conflict updates. As a result, the patch fails the benchmark test suite.

In contrast, when the repair model is provided with repair suggestions, \textit{django\_\_django-17051} can be correctly fixed. The suggester analyzes the call relationship between \textit{\_batched\_insert()} and \textit{\_insert()} and explicitly identifies that, beyond modifying the conditional statement, it is also essential to pass the conflict-handling parameters in the \textit{\_insert()} call. By introducing the three parameters \textit{on\_conflict, update\_fields, unique\_fields}, the patch achieves full compatibility with the \texttt{ON CONFLICT DO UPDATE} semantics, enabling the system to correctly return primary key IDs when \textit{update\_conflicts=True} is specified, thereby completely resolving the issue.

The motivation example demonstrates that the suggester plays a crucial role not only in producing syntactically valid repair suggestions but also in injecting semantic context awareness into the repair process. Without repair suggestions, the repair model tends to generate superficially correct but structurally incomplete patches. With the suggester enabled, the repair model can reason across function-level dependencies and produce structurally complete and semantically correct repairs. In our work, we introduce an additional suggester module between localization and repair, enabling the repair model to better understand the bug and guide the fixing process.

\section{Methodology}
\subsection{Problem Statement}

Assuming \( B \) is a bug to be fixed and \( C \) is the codebase in which it resides, the existing localize-then-fix repair process can be defined as follows.

\find{
\begin{definition}[The localize-then-fix Repair Paradigm]
\label{def:locate_fix}
The localize-then-fix process consists of two stages. First, the localization model $\alpha$ identifies the locations $L$ of bug $B$ based on the bug description and the codebase $C$. Then, the repair model $\beta$ generates a patch $P$ for the bug $B$ using $B$, $C$, and the identified locations $L$.

\begin{equation}
\pi_{\theta}(L, P \mid C, B) = 
\underbrace{\pi_{\alpha}(L \mid C, B)}_{\text{Localization}} \cdot 
\underbrace{\pi_{\beta}(P \mid C, B, L)}_{\text{Fixing}}
\end{equation}

\end{definition}
}

Building upon the localize-then-fix process, \toolname{} further introduces a suggest phase between localization and repair to generate repair suggestions, enabling the repair model to better understand the bug and refine the repair objective. The locate-suggest-fix process is defined as follows.

\find{
\begin{definition}[The locate-suggest-fix Repair Paradigm]
\label{def:locate_suggest_fix}
The key difference between locate-suggest-fix and localize-then-fix lies in the introduction of a suggester parameterized by $\gamma$ after bug localization. Specifically, after the location $L$ is identified, the suggester $\gamma$ generates a repair suggestion $R$ based on the location $L$, the bug $B$, and the codebase $C$. This suggestion $R$ is then used together with $L$ as input in the subsequent repair phase.
\begin{equation}
\pi_{\theta}(L, R, P \mid C, B) = 
\underbrace{\pi_{\alpha}(L \mid C, B)}_{\text{localizer}} \cdot 
\underbrace{\pi_{\gamma}(R \mid C, B, L)}_{\text{suggester}} \cdot 
\underbrace{\pi_{\beta}(P \mid C, B, L, R)}_{\text{fixer}}
\end{equation}
\end{definition}
}

\subsection{Overview}
\begin{figure}[t]
    \centering
    \includegraphics[width=\textwidth]{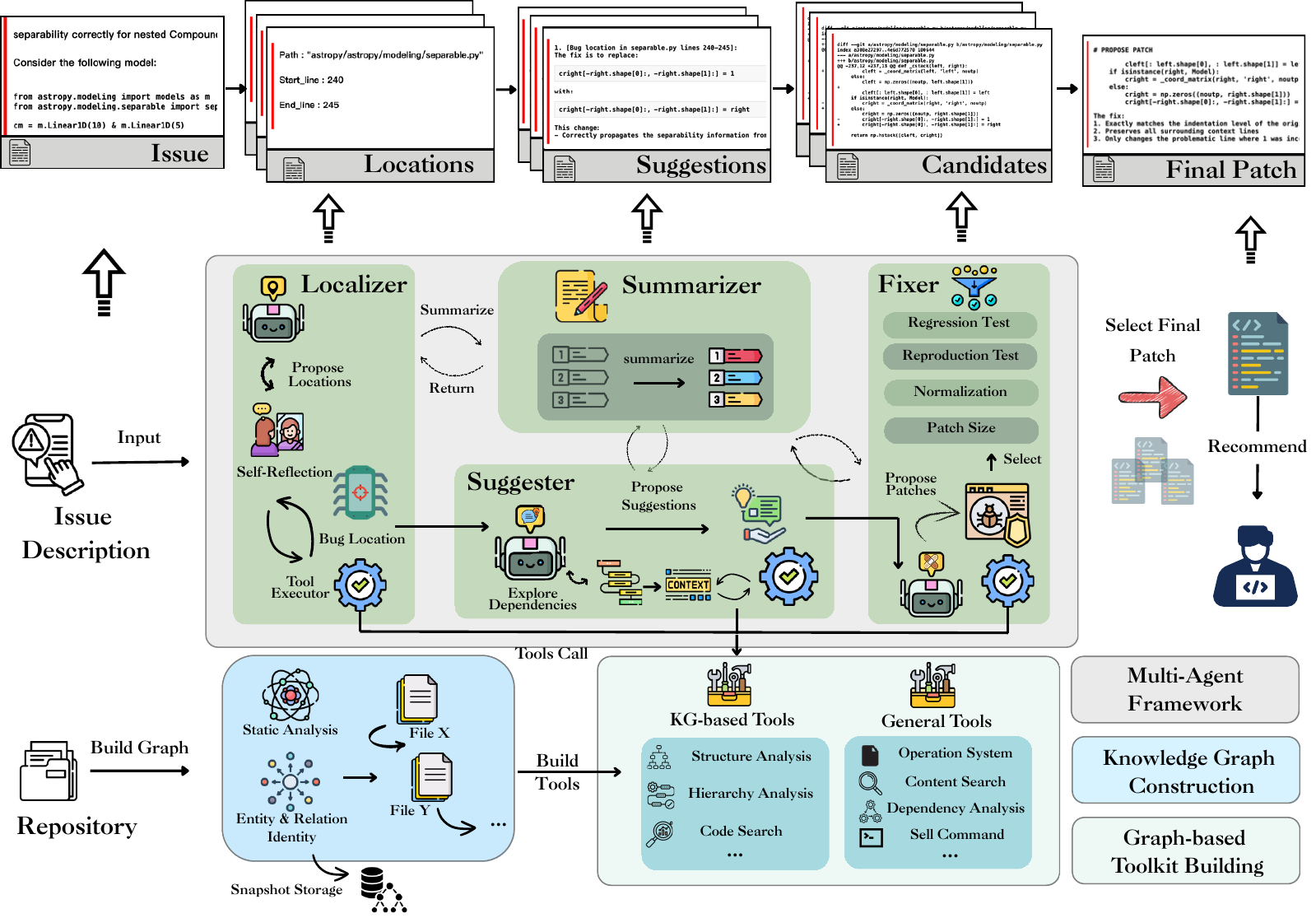}
    \caption{Overview of \toolname{}.}
    \Description{A diagram showing the workflow of \toolname{}.}
    \label{fig:workflow}
\end{figure}

The overview of \toolname{} is shown in Figure~\ref{fig:workflow}.
\toolname{} is a suggestion-guided multi-agent software repair framework that takes a GitHub issue and the corresponding codebase as inputs, and outputs a validated patch.
Specifically, \toolname{} is composed of three core components: (1) Knowledge Graph: a self-implemented repository graph consisting of three types of entities and seven types of relationships. This graph serves as the knowledge base capturing both the structural and semantic information of the repository. (2) Knowledge Graph Toolkit: a complex, knowledge-graph-driven toolkit that enables agents to perform accurate and efficient context retrieval across diverse repair scenarios. (3) Multi-Agent Framework: composed of the localizer, suggester, and fixer agents, which collaborate under a locate–suggest–fix paradigm to iteratively identify, analyze, and repair repository-level software bugs. Next, we introduce the core components of \toolname{} in sequence.

\textbf{Knowledge Graph Construction.} First, \toolname{} constructs a structured knowledge graph based on the hierarchical organization of the target repository. It encodes code entities along with their structural and semantic relationships, serving as a knowledge base to support repository-level code understanding and contextual retrieval for the Multi-Agent system.

\textbf{KG-based Toolkit Implementation.} Second, to enable the agents to efficiently utilize the knowledge graph for context retrieval, \toolname{} builds a retrieval toolkit encompassing multiple querying mechanisms tailored to common code retrieval scenarios. Agents can invoke different tools within the toolkit depending on the task context, while the tools themselves provide additional contextual cues and guidance to optimize the retrieval process.

\textbf{Multi-Agent Framework.} \toolname{} adopts a three-stage agent framework consisting of Locate, Suggest, and Fix phases, each handled by a specialized agent. The localizer retrieves relevant contextual information from the knowledge graph and proposes candidate bug locations. The suggester performs secondary retrieval based on the identified locations, refines the repair objectives, and generates repair suggestions. The fixer synthesizes patches according to the localization and suggestions, then performs patch ranking and selection to determine the final repair output.

In the following sections, we will present the design rationale and technical implementation of these three components.

\subsection{Knowledge Graph Construction}
To facilitate repository-level code retrieval and reasoning for LLMs, we construct a structured, repair-oriented knowledge graph based on static code analysis.
To this end, we primarily use AST parsing to extract code entities and structural anchors, and then augment them with cross-file dependency relations that are directly useful for repository-level repair.
Together with a custom-designed retrieval toolkit, this knowledge graph enables the agent to navigate large software repositories more effectively and to retrieve semantically relevant context for downstream reasoning and patch generation.

Particularly, the construction process takes the complete repository as input, including both file structures and contents, and produces a graph composed of nodes and edges.
The resulting graph can interact with a Neo4J database~\cite{neo4j}, providing optional persistent storage for efficient querying and visualization.
In particular, the knowledge graph is built through the following four stages:

(1) \textit{Tag Extraction:}
For each file \( F_i \) in the project, we employ abstract syntax tree analysis to extract raw tag information with Tree-sitter.
If tag parsing fails to retrieve corresponding definition details, we additionally use Pygments to locate and supplement missing definitions, ensuring that all entities are fully captured.
At this stage, ASTs are only used as the basis for entity extraction; they do not constitute the final graph structure.

(2) \textit{File Structure Analysis:}
We then parse each file \( F_i \) to obtain its hierarchical structure, capturing file-level and intra-file structural relations such as class containment, method containment, and variable ownership.
These relations provide the structural backbone of the graph.
The relationships covered in our knowledge graph are shown in Table~\ref{table:relationship}.

(3) \textit{Entity and Relation Construction:}
As detailed in Table~\ref{table:attribute}, we define three categories of entities, including classes, methods, and variables.
Based on the tag information obtained in Step 1, we build indexed nodes for each entity through pointer analysis.
We then derive multiple dependency relations from the structural information obtained in Step 2 and further enrich them with semantic dependency relations across files and modules.
Unlike a pure AST, which is strictly confined to syntactic hierarchy within a single file, our construction combines Tree-sitter tokens with static analysis to connect usage sites (\eg, function calls) to their corresponding definition sites across the repository.
These cross-file relations transform isolated hierarchical trees into an interconnected semantic network, enabling the agent to trace dependency propagation and retrieve repair-relevant context beyond local syntax.

(4) \textit{Graph Representation:}
Finally, the knowledge graph is represented as a set of triples \((h, t, r)\), where \(h\) and \(t\) denote entities (represented as nodes) and \(r\) represents the relation between them (represented as edges).
This representation enables efficient graph retrieval and semantic reasoning during context retrieval.

It is worth noting that our goal is not to encode all possible forms of program semantics, but to construct a task-oriented knowledge graph that emphasizes relations most actionable for repository-level repair, such as containment, invocation, symbol reference, inheritance, and import-related dependencies.
This design choice distinguishes our graph from heavyweight code property graph tools such as Joern, which integrate AST, control-flow, and program dependence graphs for comprehensive static analysis.

\begin{table*}[t]
    \centering
    \footnotesize
    \caption{The relationships between entities defined in the knowledge graph.}
    \resizebox{\textwidth}{!}{
        \begin{tabular}{llll}
            \toprule
            \textbf{Relationships} & \textbf{Source Node} & \textbf{Target Node} & \textbf{Description} \\
            \midrule
            INHERITS      & Class    & Class    & Inheritance from child class to parent class. \\
            HAS\_METHOD   & Class    & Method   & A class contains a method. \\
            BELONGS\_TO   & Method   & Class    & A method belongs to a class. \\
            HAS\_VARIABLE & Class    & Variable & A class contains a variable or constant. \\
            BELONGS\_TO   & Variable & Class    & A variable belongs to a class. \\
            CALLS         & Method   & Method   & Dependency of cross-file or intra-file method invocations. \\
            REFERENCES    & Method   & Class    & Dependency linking usages to their corresponding definitions. \\
            \bottomrule
        \end{tabular}
    }
    \label{table:relationship}
\end{table*}

\begin{table*}[t]
    \centering
    \footnotesize
    \caption{The attributes of the entities defined in the knowledge graph.}
    \begin{tabular}{l|l|l}
        \toprule
        \textbf{Attributes} & \textbf{Data Type} & \textbf{Description} \\
        \midrule
        \multicolumn{3}{l}{\textbf{Class}} \\
        \midrule
        fully qualified name & String      & Fully qualified name of the class \\
        name                 & String      & Name of the class \\
        absolute path        & String      & Absolute path of the class \\
        start line           & Integer     & Start line number of the class \\
        end line             & Integer     & End line number of the class \\
        content              & String      & Source code of the class \\
        class type           & String      & Class type: normal or inner \\
        parent class         & String      & Fully qualified name of the parent class \\
        \midrule
        \multicolumn{3}{l}{\textbf{Method}} \\
        \midrule
        fully qualified name & String      & Fully qualified name of the method \\
        name                 & String      & Name of the method \\
        absolute path        & String      & Absolute path of the method \\
        start line           & Integer     & Start line number of the method \\
        end line             & Integer     & End line number of the method \\
        content              & String      & Source code of the method \\
        params               & Json String & Parameter list of the method \\
        modifiers            & Json String & Modifiers of the method (decorator and access level) \\
        signature            & String      & Signature of the method \\
        type                 & String      & Constructor or normal \\
        \midrule
        \multicolumn{3}{l}{\textbf{Variable}} \\
        \midrule
        fully qualified name & String      & Fully qualified name of the variable \\
        name                 & String      & Name of the variable \\
        absolute path        & String      & Absolute path of the variable \\
        start line           & Integer     & Start line number of the variable \\
        end line             & Integer     & End line number of the variable \\
        content              & String      & Source code of the variable \\
        modifiers            & Json String & Modifiers of the variable \\
        data type            & String      & Data type of the variable \\
        \bottomrule
    \end{tabular}
    \label{table:attribute}
\end{table*}

\subsection{KG-based Toolkit Implementation}
Building upon the knowledge graph, we design a comprehensive retrieval toolkit that facilitates efficient interaction between \toolname{} and the whole repository.
The toolkit provides \toolname{} with precise and context-aware retrieval utilities that are preconfigured for multiple scenarios. A detailed description of each tool and its functionality is summarized in Table \ref{table:toolkit}, and we omit exhaustive listings here due to space constraints.

The design of the toolkit is motivated by the natural workflow of human developers during real-world debugging. 
In practice, developers rarely interact with a repository through a single monolithic operation. 
Instead, they usually begin by inspecting the overall code structure, then search for relevant keywords or symptoms, next analyze the definitions, usages, and relations of key entities, and finally inspect concrete file content or system states before implementing modifications. 
To simulate this debugging process, we intentionally organize the toolkit into four categories, allowing the agent to navigate the repository in a structured and human-like manner. 
Moreover, we deliberately design the tools as lightweight and composable primitives, rather than embedding overly complex reasoning into a single tool. 
This design preserves the flexibility of the agent, which serves as the reasoning engine and determines which tools to invoke, in what order, and how to combine their outputs.

(1) \textit{Code Structure Analysis Tools.}
These tools focus on extracting and analyzing the structural information of the repository, such as file hierarchies and dependency relations between entities. For example, \textit{analyze\_file\_structure} tool conducts a complete overview of a Python file, listing all classes and methods with their names, full qualified names, and parameters. It serves as an essential starting point for understanding file architecture.

(2) \textit{Entity Analysis Tools.}
These tools handle retrievals related to code entities, including classes, methods, and variables (\eg \textit{find\_class\_constructor} tool extracts class constructor with full implementation). We further extend this category with import-analysis tools to capture inter-file dependencies introduced by module imports (\eg \textit{show\_file\_imports} tool extracts all import statements from a Python file).

(3) \textit{Content Search Tools.}
These tools support contextual retrieval based on specific content or keywords within the codebase, helping agents locate relevant entities. For instance, \textit{search\_code\_with\_context} tool searches for keywords in Python files with 3 lines before and after each match.

(4) \textit{File System Tools.} 
These tools enable system-level interactions, including directory traversal , file reading and command-line execution.

\begin{table*}[t]
    \centering
    \caption{The KG-based toolkit used by \toolname{}.}
    \resizebox{\linewidth}{!}{
        \begin{tabular}{ll}
            \toprule
            \textbf{Tools} & \textbf{Description} \\
            \midrule
            \multicolumn{2}{l}{\textbf{Code Structure Analysis Tools}} \\
            \midrule
            analyze\_file\_structure
                & Get a complete overview of a Python file \\
            get\_code\_relationships
                & Discover how any code entity (method, class, or variable) connects to other code \\
            find\_methods\_by\_name
                & Locate all methods with a specific name across the entire project with simplified relationship analysis. \\
            \midrule
            \multicolumn{2}{l}{\textbf{Entity Analysis Tools}} \\
            \midrule
            extract\_complete\_method
                & Extract full method implementation with automatic relationship analysis. \\
            find\_class\_constructor
                & Locate and extract class constructor (\_\_init\_\_ method) with full implementation. \\
            list\_class\_attributes
                & Get all field variables and attributes defined in a class, including their data types and content. \\
            find\_variable\_usage
                & Search for variable usage in a specific file, showing all occurrences with line numbers and context. \\
            find\_all\_variables\_named
                & Find all variables with a specific name across the entire project. \\
            show\_file\_imports
                & Extract all import statements from a Python file. \\
            \midrule
            \multicolumn{2}{l}{\textbf{Content Search Tools}} \\
            \midrule
            search\_code\_with\_context
                & Search for keywords in Python files with surrounding code context (3 lines before and after each match). \\
            find\_files\_containing
                & Find all files that contain specific keywords in their content or filename. \\
            \midrule
            \multicolumn{2}{l}{\textbf{File System Tools}} \\
            \midrule
            explore\_directory
                & List directories and files in a given path. \\
            read\_file\_lines
                & Read specific line ranges from files with line numbers. Maximum 50 lines per call. \\
            execute\_shell\_command\_with\_validation
                & Execute read-only shell commands for system inspection. \\
            \bottomrule
        \end{tabular}
    }
    \label{table:toolkit}
\end{table*}

\subsection{Multi-Agent Framework}
To enable LLMs to autonomously leverage knowledge graphs and external tools, we design three dedicated agents inspired by the three-stage process of human debugging: the fault localizer, repair suggester, and issue fixer. 
Each agent is responsible for a distinct stage of the repair workflow. While all three agents share the same toolkit, they use it with different stage-specific objectives.

Although these three roles could in principle be implemented as distinct phases within a single agent, we adopt a multi-agent design for two reasons.
First, it improves context management.
Repository-level repair requires extensive exploration, iterative tool invocation, and repeated retrieval of code context.
If a single agent handles localization, planning, and patch generation within one unified interaction history, the accumulated tool traces and intermediate reasoning can quickly overwhelm the context window.
As a result, the information most relevant to the final repair step may be diluted.
By separating the workflow into the localizer, suggester, and fixer, each agent consumes only the condensed outputs relevant to its own subtask rather than the entire raw history.
Second, it enforces clearer task boundaries.
Localization, repair planning, and patch synthesis are related but fundamentally different reasoning tasks.
In a monolithic agent, these tasks can easily become entangled, causing the model to prematurely jump from partial localization signals to patch generation.
By assigning them to different agents with specialized prompts and responsibilities, \toolname{} better reflects human debugging practice and improves cross-stage coordination.

\subsubsection{Fault Localizer Agent}  
The localizer Agent is responsible for identifying potential bug locations using the provided issue description and repository information, supported by the available tools.
To achieve a comprehensive understanding of complex GitHub projects and accurate localization of buggy instances, the localizer is designed based on the ReAct~\cite{react} framework and performs an observe-think-act pattern and is equipped with a specialized toolkit.
It is worth noting that the contribution of the localizer is not to propose a standalone new fault localization technique, but to design localization as a non-isolated component within the \textit{locate-suggest-fix} framework, supported by the knowledge graph and specialized tools, and tightly coupled with downstream agents.

Specifically, in the Locate stage, the localizer actively leverages the bug description provided in the issue to retrieve the constructed knowledge graph through a prebuilt toolkit. It dynamically searches for contextual information, such as related code entities and their interrelationships. After each retrieval step, the localizer is guided by auxiliary assistant feedback to analyze the retrieved entities and relations, and to interpret their potential relevance to the issue. The localizer then assesses whether the current contextual information is sufficient for bug localization. If not, it iteratively invokes additional tools within the retrieval toolkit to gather more comprehensive contextual data. 
Once sufficient context is established, the localizer outputs a candidate localization set containing up to five bug locations and summarizes the accumulated insights, which are subsequently passed to the suggester Agent for suggestion generation.
Unlike prior work that treats localization as a separate stage whose outputs are forwarded directly to patch generation in a static manner, our localizer produces a set of candidate bug locations that are not taken as final repair targets, but can instead be refined into repair-oriented suggestions by the downstream suggester.

The localizer is guided by a structured prompt (shown in Listing~1) explicitly designed to ensure accurate and interpretable bug localization. The prompt instructs the model to act as a bug localization specialist that understands the issue solely through its problem description, which serves as the single source of truth. 
It requires the agent to analyze the described symptoms, infer potential root causes, and output a set of up to five locations where the bug may reside.
Each location must be expressed in a line-range format (\eg, line 42–47) and must not overlap with others. The prompt enforces minimal perfect coverage, ensuring that all buggy regions (including deleted or modified lines) are captured while avoiding redundancy.
To guarantee clarity and consistency, the prompt further defines explicit output schemas, completion signals, and structural tags (\eg, <locations>, <format>, <scope>, <constraint>), which enable deterministic parsing by downstream agents. Additionally, it emphasizes the logical or functional relationships among the identified locations, guiding the Localizer to reason across related components such as caller–callee pairs or shared-state interactions.
This carefully designed prompt transforms the localization process into a structured reasoning task, encouraging the model to produce precise, consistent, and context-aware localization outputs suitable for subsequent suggestion and fix stages.

\begin{promptbox}{Fault Localizer Agent Prompt}
\label{prompt:localizer}
Act as the localizer to understand bugs by analyzing problem descriptions, identifying root causes, and locating specific methods or line ranges in the codebase where bugs can be fixed. \newline

$<output\_requirements>$

$<locations>$

$<location~count>$up to 5 precise and interrelated locations$<count>$

$<format>$range of line numbers (\eg, line 42–47)$<format>$

$<scope>$single bug, possibly manifesting in multiple places$<scope>$

$<relationship>$logically or functionally connected$<relationship>$

$<constraint>$provided ranges must not overlap with each other $<constraint>$

$<locations>$ \newline

$<completion\_signal>$
...
$<completion\_signal>$

$<output\_requirements>$\newline

$<guidelines>$

$<truth\_source>$
The problem description is your sole source of truth.
Base your entire investigation and reasoning on it.
$<truth\_source>$

$<coverage\_requirement>$
Provide minimal perfect coverage of vulnerable code locations,
including any deleted lines.
$<coverage\_requirement>$

$<guidelines>$
\end{promptbox}

\subsubsection{Repair suggester Agent}  
The suggester Agent is responsible for retrieving additional contextual information related to the identified locations and generating feasible repair suggestions. These suggestions serve to direct the fixer Agent’s attention toward more valuable and relevant context, effectively guiding the repair process while mitigating the influence of irrelevant information.
It is worth noting that the suggester is not designed to explicitly re-localize, validate, or replace the bug locations proposed by the localizer, since such behavior would overlap with the role of the localizer. Instead, the suggester focuses on expanding the context around the candidate locations and refining these preliminary localization results into repair-oriented suggestions.

In the Suggest stage, the suggester Agent builds upon the localization results produced by the previous tools to provide targeted repair suggestions for the subsequent fixing process. 
Specifically, starting from each identified location, the suggester performs context expansion guided by the issue description until it determines that the retrieved information is sufficient to adequately understand both the root cause of the bug and potential repair directions.
Consistent with prior agent-based and ReAct-style studies~\cite{autocoderover,react}, this stopping decision is made autonomously by the model based on the current context and interaction history, rather than by a separate stopping criterion.
For each candidate location, the suggester then generates specific repair suggestions and produces an updated context summarization, which are passed to the fixer Agent for execution. 
Although the suggester does not explicitly rewrite the localizer's outputs, the secondary retrieval and context expansion process may implicitly sharpen, supplement, or correct the repair-relevant locations reflected in the generated suggestions.
By conducting this secondary exploration based on the localizer’s preliminary findings, the suggester refines the repair objective and reduces unnecessary search divergence, thereby allowing the fixer to focus on more accurate and efficient patch generation.

\begin{promptbox}{Repair Suggester Agent Prompt}
\label{prompt:suggester}
Act as the suggester, collecting relevant context and providing precise, actionable repair suggestions for each bug location provided by the localizer.\newline

$<location\_analysis>$

$<interconnection>$
These locations are functionally interconnected.
Analyze how they relate to each other and together contribute to the bug.
$<interconnection>$ \newline

$<coordination>$
Suggestions should reflect this interconnectedness.
While you may propose fixes for each location separately,
they must work in coordination to fully resolve the bug.
$<coordination>$ \newline

$<assumptions>$
Do not question or validate the correctness of bug locations.
Assume all provided locations are valid.
Focus solely on understanding them as a whole and offering feasible, context-aware fixes.
$<assumptions>$

$<location\_analysis>$ \newline

$<framework\_preservation>$

$<heritage\_check>$Maintain framework design patterns$<heritage\_check>$

$<context\_integrity>$Preserve error location tracking$<context\_integrity>$

$<api\_compliance>$Use framework-approved interfaces$<api\_compliance>$

$<framework\_preservation>$ \newline

$<output\_format>$

$<feedback\_handling>$
...
$<feedback\_handling>$

$<completion\_header>$
...
$<completion\_header>$

$<structure\_example>$
...
$<structure\_example>$

$<output\_format>$
\end{promptbox}

The suggester operates under a structured prompt (shown in Listing~2) designed to guide it in performing extra context retrieval and generating coherent, framework-compliant repair suggestions. 
The prompt positions the model as a context-aware repair planner, instructing it to analyze the logically interconnected bug locations produced by the localizer and to infer how these regions collectively contribute to the underlying defect. 
Rather than validating or modifying the provided locations, the suggester is required to treat them as correct and focus exclusively on understanding their interactions and generating actionable repair strategies.
To ensure contextual consistency, the prompt enforces several key principles: (1) interconnection analysis: requiring the suggester to reason about how different code locations are functionally related; (2) coordination constraints: ensuring that proposed fixes across locations remain logically and functionally aligned; and (3) framework preservation: mandating compliance with project-specific design rules such as maintaining framework patterns, ensuring context integrity, and adhering to tools provided.
The output schema explicitly specifies structured tags (\eg <location\_analysis>, <framework\_preservation>, <output\_format>) and completion signals to maintain syntactic uniformity and enable downstream parsing by the fixer. This design compels the suggester to transform loosely defined context exploration into a structured reasoning process, producing targeted, framework-consistent repair suggestions that bridge the gap between localization and patch synthesis.

\subsubsection{Issue Fixer Agent}
The issue fixer agent is responsible for generating candidate patches from suspicious code snippets and repair suggestions provided by upstream agents, and for validating and selecting the final patch that resolves the bug. Specifically, the fixer Agent is guided to conduct a structured reasoning process over the provided contextual information and repair suggestions. 
First, the fixer analyzes the root cause of the bug based on the retrieved repository context and the suggestions generated by the suggester. When necessary, the fixer is allowed to retrieve additional contextual or dependency information from the knowledge graph or the toolkit to support its reasoning process. 
Unlike the root cause analysis performed by the Suggester, which focuses on refining localization results into repair-oriented plans, the Fixer’s analysis focuses on grounding the suggested repair in the concrete repository state and verifying how it should be instantiated as executable code changes. 
Next, the fixer is explicitly instructed to explain why the proposed repair is expected to resolve the bug described in the issue. This explanatory step enforces interpretability and helps ensure that the generated patch aligns semantically with the intended fix rather than being coincidentally valid.
Finally, the fixer generates an executable patch that can be directly applied to the repository for validation through regression and reproduction testing.
In particular, the fixer Agent synthesizes the localization results, repair suggestions, and contextual information to generate multiple patch candidates for each identified location.

After patch generation, inspired by Agentless~\cite{agentless}, \toolname{} adopts a three-phase validation pipeline to ensure the functional correctness and semantic reliability of the generated patches.
First, the fixer executes regression tests on each patch \( P_i \), recording the number of passed tests \( T_{regression} \). Based on these results, the patches are ranked, and the top-performing set \( P_{regression} \) is selected.
Next, \toolname{} employs a separate LLM to automatically construct a reproduction test according to the issue description, aiming to replicate the reported bug. The fixer then executes these reproduction tests on each patch in \( P_{regression} \), records the number of successful passes \( T_{reproduction} \), and re-ranks the results to obtain the top-performing set \( P_{reproduction} \).
Finally, we normalize the remaining patches and apply a majority voting process, selecting the one with the most consistent semantics as the final patch. This multi-stage validation strategy enables \toolname{} to produce patches that are both functionally correct and semantically aligned with the intended repair objective.

Unlike the open-ended exploration of upstream agents, the fixer's retrieval behavior is strictly local and closed. It does not search for new bug locations or alternative repair strategies. Instead, its retrieval is triggered solely by the need for precise code-level alignment right before generating the final patch. By invoking restricted tools (\eg, read\_file\_lines) to verify specific physical details—such as exact line numbers, variable names, and indentations—the fixer mitigates the risk of hallucinating physical offsets when translating high-level suggestions into executable code. This localized verification ensures that the generated patch is physically compatible with the repository, serving as a micro-level execution step that perfectly complements the macro-level decision flow.

The fixer operates under a task-oriented prompt (shown in Listing~3) designed to synthesize, justify, and implement executable patches based on the contextual information and repair suggestions provided by upstream agents. The prompt positions the model as a code repair executor that must reason over the aggregated evidence from the localizer and suggester, analyze the root cause of the defect, and produce a coherent and verifiable patch that resolves the described issue.
The prompt enforces a systematic workflow comprising three key stages:
(1) \textit{Pre-implementation analysis}: The fixer is instructed to analyze the defect holistically using the provided context and suggestions, explain the rationale behind its proposed solution, and confirm that the resulting patch is expected to pass the corresponding test suite.
(2) \textit{System-aware repair generation}: When multiple interconnected locations are involved, the fixer must treat them as a unified system, ensuring that each modification contributes to a consistent and functionally complete repair across all affected code regions.
(3) \textit{Patch quality assurance}: The prompt specifies explicit requirements for output structure, code quality, and framework compatibility (\eg, reuse of existing functions, minimal new definitions, and adherence to project conventions).
The prompt includes structured fields such as <system\_approach>, <pre\_implementation\_analysis>, and <patch\_requirements> to ensure well-formed outputs and to facilitate downstream automated validation and ranking. By incorporating explicit reasoning and verification steps, the fixer prompt transforms code generation into a transparent, explainable process-bridging high-level repair intent with concrete, executable solutions that meet both functional and structural correctness criteria.

\begin{promptbox}{Issue Fixer Agent Prompt}
\label{prompt:fixer}
Implement a complete and correct bug fix based on the localizer's identified code locations and the suggester's proposed suggestions.
The project is written in Python, and bugs may occur at any granularity, \eg, from a single variable to an entire module. 

$<system\_approach>$

$<interconnected\_locations>$

There may be multiple interconnected locations to fix.
Treat them as a whole system: each fix must work together to fully resolve the bug.

$<interconnected\_locations>$ \newline

$<pre\_implementation\_analysis>$

Before presenting your fix, you should:

- Analyze the root cause based on context and suggestions

- Briefly explain why your solution addresses the issue holistically

- Confirm that applying your patch causes the test suite to pass

- Reuse existing code when possible; only define new functions or variables if necessary

$<pre\_implementation\_analysis>$

$<system\_approach>$ 

$<patch\_requirements>$

$<location\_matching>$
...
$<location\_matching>$

$<output\_format>$
...
$<output\_format>$

$<code\_quality\_standards>$
...
$<code\_quality\_standards>$

$<framework\_compatibility>$
...
$<framework\_compatibility>$

$<patch\_requirements>$
\end{promptbox}

\subsubsection{Dynamic Memory Summarization}

Since the performance of LLMs is often sensitive to context length, excessively long conversational histories can lead to context dilution and hallucination, significantly impairing the reliability of \toolname{}. 
To mitigate this issue, we introduce a summarization mechanism that dynamically compresses historical interactions while preserving essential contextual information for reasoning and retrieval.
Specifically, \toolname{} maintains a message window for each agent. When the accumulated tokens within the queue exceed a predefined threshold $t_{max}$, a dedicated summarization LLM is invoked to condense the current conversation history. The generated summary is then appended as part of the active context to guide subsequent exploration and reasoning. After summarization, only the most recent $m_{remain}$ messages are retained, and older messages are discarded. When discarding outdated messages, \toolname{} ensures the structural integrity of tool interactions. Specifically, if a tool call message is subject to deletion while its corresponding tool response must be retained, the associated tool call is preserved as well. This guarantees the completeness of tool-related message pairs and prevents information loss that could otherwise disrupt reasoning consistency or contextual grounding during subsequent agent operations.
Furthermore, when a summary already exists and the message window reaches its capacity, the summarizer is instructed to integrate the previous summaries into the newly generated one, thereby ensuring contextual continuity and minimizing cumulative information loss. This mechanism strikes a balance between long-context retention and memory efficiency, enabling \toolname{} to maintain coherent reasoning and stable performance in large-scale repository-level repair tasks.

\section{Experimental Setup}
\subsection{Research Questions}
We evaluate \toolname{} on the following research questions:

\textbf{RQ1:} How does \toolname{} perform compared to existing baselines?

\textbf{RQ2:} How do each component contribute to the performance of \toolname{}?

\textbf{RQ3:} How does \toolname{} perform with different base models?

\textbf{RQ4:} How does \toolname{} perform when extended to vulnerability repair tasks?

\subsection{Benchmark}
We evaluate \toolname{} on the repository-level software repair dataset SWE-Bench~\cite{swe-bench}. Specifically, we select SWE-Bench-Lite, a subset of SWE-Bench, as the primary benchmark because it has been widely evaluated and is more cost-efficient for experimentation. SWE-Bench-Lite collects and filters issues from twelve actively maintained GitHub python projects, resulting in a total of 300 real-world, complete, and reproducible bug instances. Each bug instance includes a full snapshot of the project repository, an executable test suite, and a natural-language issue description.

In recent years, SWE-Bench-Lite has become widely adopted among researchers due to its realism and complexity. It features large-scale software projects, with some repositories containing tens of thousands of lines of code, and buggy instances requiring patches that often span multiple locations across different files to repair. Consequently, APR tools must traverse extensive codebases, sometimes across hundreds of files, to identify the root causes of bugs and generate effective patches. This makes SWE-Bench-Lite a rigorous testbed for evaluating repository-level comprehension and cross-file reasoning capabilities.

\subsection{Baselines}
To address our RQs, we compare \toolname{} with state-of-the-art repair systems on the SWE-Bench-Lite leaderboard as baselines, including AutoCodeRover~\cite{autocoderover}, SWE-Agent~\cite{sweAgent}, DARS-Agent~\cite{dars}, OpenHands~\cite{openhandsArticle}, Agentless~\cite{agentless}, KGCompass~\cite{KGCompass}, ExpeRepair~\cite{experepair}.
DARS-Agent proposes a novel inference time compute scaling approach for coding agents. Lingxi introduces an agent manager to manage the software repair process and a supervisor to oversee the entire workflow of solving the issue, and decide which agent to route to depending on the current progress of the issue. OpenHands proposes a platform for the development of powerful and flexible AI agents, allowing for the implementation of new agents, the utilization of various LLMs, safe interaction with sandboxed environments for code execution, and the incorporation of evaluation benchmarks.
Table~\ref{tab:related} summarizes the main differences between \toolname{} and representative prior work with publicly available pre-prints or published papers.
The last column indicates whether an approach introduces an explicit intermediate stage between localization and patch generation.

\begin{table}[t]
  \centering
  \footnotesize
  \caption{Comparison between \toolname{} and state-of-the-art approaches}
    \begin{tabular}{cccccc}
    \toprule
    \textbf{Approach} & \textbf{Time} & \textbf{Publisher} & \textbf{Paradigm} & \textbf{KG} & \textbf{Intermediate Stage} \\
    \midrule
    AutoCodeRover~\cite{autocoderover} & 2024 & ISSTA & Multi-agent & ×     & × \\
    SWE-Agent~\cite{sweAgent} & 2024 & NeurIPS   & Single-agent & ×     & × \\
    DARS-Agent~\cite{dars} & 2025 & ACL   & Single-agent & ×     & × \\
    OpenHands~\cite{openhandsArticle} & 2025 & ICLR  & Multi-agent & ×     & × \\
    Agentless~\cite{agentless} & 2025 & FSE   & Workflow & ×     & × \\
    KGCompass~\cite{KGCompass} & 2025 & arXiv & Workflow & \checkmark  & × \\
    ExpeRepair~\cite{experepair} & 2025 & arXiv &  Multi-agent & ×     & × \\
    \midrule
    \toolname{} & N/A   & N/A   & Multi-agent & \checkmark     & \checkmark \\
    \bottomrule
    \end{tabular}%
  \label{tab:related}%
\end{table}%

\subsection{Evaluation Metrics}
Following the standard practice of previous work~\cite{KGCompass, agentless}, we adopt four metrics to assess the performance of \toolname{} on SWE-Bench-Lite, including \% Resolved, File Acc., Func Acc., and Avg Cost per bug. 
\% Resolved measures the proportion of instances whose final patch successfully passes all benchmark tests. 
File Acc. evaluates file-level localization accuracy, computed as the average Jaccard similarity between predicted and ground-truth file sets. 
Func Acc. measures function-level localization accuracy. 
Avg Cost per bug represents the average token consumption required to resolve each bug, reflecting the computational efficiency of \toolname{}.
For each buggy input instance $i$, we use the following equation to compute the accuracy of file-level localization.

\begin{equation}
a_i^{\text{file}} = \frac{ \left| F_i^{\star} \cap \hat{F}_i \right| }{ \left| F_i^{\star} \cup \hat{F}_i \right| }
\end{equation}

Where $ F_i^{\star} $ denotes the ground-truth set of files to be modified in instance $ i $, and $ \hat{F}_i $ denotes the predicted set of files for instance $ i $. For function-level localization, since Java syntax typically associates one file with one class, we do not distinguish between classes and files. Thus, function-level localization accuracy can be calculated by:
\begin{equation}
a_i^{\text{func}} = \frac{ \left| E_i^{\star} \cap \hat{E}_i \right| }{ \left| E_i^{\star} \cup \hat{E}_i \right| }
\end{equation}

Where $ E_i^{\star} $ denotes the ground-truth set of files/classes requiring edits in instance $ i $, and $ \hat{E}_i $ denotes the predicted set of files/classes. Based on file-level and function-level localization accuracy, the average accuracy of localization over a specific set of bugs $S$ can be further calculated by:
\begin{equation}
\text{FileAcc} = \frac{1}{|S|} \sum_{i \in S} a_i^{\text{file}}, 
\qquad
\text{FuncAcc} = \frac{1}{|S|} \sum_{i \in S} a_i^{\text{func}}
\end{equation}

These metrics collectively provide a comprehensive and objective evaluation of \toolname{}’s localization accuracy, repair effectiveness, and computational efficiency.

\subsection{Implementation Details}
All of our approaches are built on the LangChain~\cite{langchain} and LangGraph~\cite{langgraph} frameworks.
We primarily evaluate \toolname{} using Claude-3.5-Sonnet~\cite{claude} as the base model. To assess the model-agnostic capability of the framework, we further evaluate \toolname{} with two alternative LLMs: DeepSeek-V3~\cite{deepseekv3} and Qwen3-235B-A22B~\cite{yang2025qwen3}. All models are used with their default hyperparameters, and no explicit limit is imposed on output tokens. 
Following prior work~\cite{agentless}, for each issue, we sample four localization sets, and each set contains up to five candidate bug locations.
For each location, patches are sampled once with temperature 0 and nine times with temperature 0.8 to enhance patch diversity.
For knowledge graph construction, we employ Tree-sitter~\cite{tree-sitter}, Python AST~\cite{python-ast}, and Pygments~\cite{pygments} to extract syntactic and structural information from the source code, forming the foundation for repository-level context retrieval.

\section{Results and Analysis}
\subsection{RQ1: Comparison With State-of-the-Art Methods}
\subsubsection{Design}
RQ1 aims to evaluate the repository-level software repair capability of \toolname{} and compare it against state-of-the-art methods on the SWE-Bench-Lite benchmark. All baseline methods employ models from the Claude family as their base LLMs (including Claude-4, Claude-3.7, and Claude-3.5).
In this research question, we use Claude-3.5 as the base model for \toolname{} and conduct comparisons across four metrics: proportion of instances resolved(\% Resolved), file-level localization accuracy (File Acc.), function-level localization accuracy (Func Acc.), and Avg Cost per bug (cost).
Additionally, we perform a cross-analysis among all Claude-3.5-based approaches to determine whether \toolname{} is capable of resolving unique buggy instances that other approaches fail to repair.

\begin{table*}[t]
    \centering
    \footnotesize
    \caption{A comparison of repair results between \toolname{} and existing baselines on SWE-Bench-Lite.}
        \begin{tabular}{cccccc}
            \toprule
            \textbf{Approaches} & \textbf{Base Model} & \textbf{\% Resolved} & \textbf{File Acc.} & \textbf{Function Acc.} & \textbf{Cost} \\
            \midrule
            ExpeRepair-v1.0 & Claude-4-Sonnet \& o4-mini       & 181 (60.3\%) & 84.7 & 52.6 & $\sim$\$2.1 \\
            Refact.ai Agent & Claude-3.7-Sonnet \& o4-mini     & 180 (60.0\%) & 80.6 & 47.0 & N/A \\
            KGCompass       & Claude-4-Sonnet                  & 175 (58.3\%) & 83.6 & 56.0 & \$0.2 \\
            SWE-Agent       & Claude-4-Sonnet                  & 170 (56.7\%) & 80.9 & 53.9 & $\sim$\$1.6 \\
            ExpeRepair-v1.0 & Claude-3.5-Sonnet \& o3-mini     & 145 (48.3\%) & \cellcolor[HTML]{FFFFFF}80.7 & 49.6 & \$2.1 \\
            SWE-Agent       & Claude-3.7-Sonnet                & 144 (48.0\%) & 79.3 & 52.2 & $\sim$\$1.6 \\
            DARS-Agent      & Claude-3.5-Sonnet \& DeepSeek-R1 & 141 (47.0\%) & 78.4 & 49.3 & \$12.24 \\
            KGCOMPASS       & Claude-3.5-Sonnet                & 137 (46.0\%) & 76.7 & 49.4 & \$0.2 \\
            Lingxi          & Claude-3.5-Sonnet                & 128 (42.7\%) & 66.1 & 39.7 & N/A \\
            OpenHands       & Claude-3.5-Sonnet                & 125 (41.7\%) & 69.3 & 43.8 & \$1.3 \\
            Agentless-1.5   & Claude-3.5-Sonnet                & 123 (41.0\%) & 79.7 & 51.8 & \$1.3 \\
            AutoCodeRover   & GPT-4o                           & 92 (30.7\%)  & 66.8 & 38.3 & \$1.3 \\
            \midrule
            \toolname{}     & Claude-3.5-Sonnet                & 154 (51.3\%) & 81.2 & 52.4 & \$1.48 \\
            \bottomrule
        \end{tabular}
    \label{table:all}
\end{table*}

\begin{table}[t]
    \centering
    \footnotesize
    \caption{Unique fixes achieved by \toolname{} and state-of-the-art approaches, all using Claude-3.5 as the base model.}
        \begin{tabular}{cccc}
            \toprule
            \textbf{Approaches} & \textbf{Base Model} & \textbf{Unique Fix by \toolname{}} & \textbf{Unique Fix by baselines} \\
            \midrule
            DARS-Agent      & Claude-3.5-Sonnet & 37 & 13 \\
            KGCompass       & Claude-3.5-Sonnet & 46 & 16 \\
            Lingxi          & Claude-3.5-Sonnet & 40 & 26 \\
            OpenHands       & Claude-3.5-Sonnet & 47 & 29 \\
            Agentless-1.5   & Claude-3.5-Sonnet & 43 & 32 \\
            \bottomrule
        \end{tabular}
    \label{tab:unique}
\end{table}

\subsubsection{Results}
Table \ref{table:all} summarizes the performance comparison between \toolname{} and ten state-of-the-art approaches on the SWE-Bench-Lite benchmark. Overall, \toolname{} demonstrates a dominant performance advantage among all approaches that use Claude-3.5 as the base model, and even surpasses several methods that rely on newer or more complex LLMs. These results provide strong evidence of \toolname{}’s effectiveness in repository-level software repair.

Specifically, \toolname{} successfully resolves 154 buggy instances, achieving a 51.3\% resolution rate, outperforming all other Claude-3.5-based baselines. Across all evaluated methods, \toolname{} ranks 5th overall, exceeding the performance of SWE-Agent, which uses Claude-3.7-Sonnet as its backbone\footnote{Our method has been displayed at the official SWE-Bench-Lite leaderboard under the name \textsc{Isea}. Details can be found at \url{https://www.swebench.com/}}.
In terms of localization accuracy, \toolname{} achieves the highest file-level accuracy (81.2\%) and function-level accuracy (52.4\%) among all Claude-3.5-based methods, ranking 3rd and 4th overall, respectively. These results validate that our knowledge-graph-based context retrieval effectively enhances bug localization performance of \toolname{}.

Table~\ref{tab:unique} presents a pairwise comparison of unique fixes between \toolname{} and five strong baselines under the same Claude-3.5-Sonnet setting, including DARS-Agent, KGCompass, Lingxi, OpenHands, and Agentless-1.5.
The results show that \toolname{} consistently resolves a substantial number of bugs that these baselines fail to fix. 
Specifically, compared with DARS-Agent, KGCompass, Lingxi, OpenHands, and Agentless-1.5, \toolname{} uniquely resolves 37, 46, 40, 47, and 43 instances, respectively, whereas these baselines uniquely fix only 13, 16, 26, 29, and 32 instances that \toolname{} misses. 
This result shows that \toolname{} not only performs better overall but also repairs buggy instances that other competitive methods fail to resolve.

In terms of computational cost, \toolname{} requires \$1.48 per instance, which is slightly higher than lightweight procedural methods such as Agentless (\$1.3), but remains competitive among agent-based frameworks. 
In general, agent-based repair methods tend to be more costly than procedural pipelines, because they require multi-turn reasoning, dynamic context retrieval, and iterative validation, whereas procedural methods usually follow a more compact and static workflow. 
For example, SWE-Agent costs about \$1.6 per instance and ExpeRepair costs about \$2.1 per instance, while \toolname{} achieves stronger same-model performance at a lower or comparable cost. This average cost is mainly attributed to the three core agents, including the Localizer (\$0.35), the Suggester (\$0.59), and the Fixer (\$0.54), while knowledge graph construction introduces no LLM API cost. Among them, the Suggester accounts for the largest share due to its additional context exploration and repair planning. Overall, these results suggest that \toolname{} maintains a favorable cost--performance trade-off despite the overhead introduced by multi-agent collaboration.

\summary{1}{\textbf{Findings:} (1) Software Repair Capability: \toolname{} achieves a 51.3\% resolution rate, outperforming all baseline approaches that use the same base model and ranking fifth overall among all evaluated methods. (2) Localization capability: \toolname{} attains 81.2\% file-level and 52.4\% function-level localization accuracy, surpassing all same-base-model baselines and ranking third and fourth, respectively, among all approaches. (3) Efficiency: \toolname{} demonstrates strong cost efficiency, requiring only \$1.48 per instance, ranking third lowest among all methods. (4) Uniqueness: \toolname{} ranks second in its ability to repair 14 unique instances that no other baseline approach can fix.
\\
\textbf{Insights:} By introducing the locate-suggest-fix paradigm, \toolname{} effectively extends the reasoning and repair boundaries of agent-based approaches, enabling more comprehensive and accurate repository-level software repair.
}

\subsection{RQ2: Ablation Study}
\subsubsection{Design}
RQ2 investigates the independent contributions of the major components in \toolname{} to repository-level software repair performance.
We conduct ablation studies by removing three key components: (1) the Suggest module, (2) the Knowledge Graph module, and (3) the Re-ranking module, resulting in three \toolname{} variants.
For the Suggest and Knowledge Graph module, we compare the variants with the full \toolname{} approach across \% Resolved, File Acc., and Func Acc. to comprehensively evaluate the impact of the Suggest and the Knowledge Graph on localization accuracy and repair effectiveness. For the Re-ranking module, we evaluate its independent contribution using only the \% Resolved metric. To this end, we conduct a progressive enhancement analysis to examine the individual effects of majority voting, regression tests, and reproduction tests. Specifically, the Re-ranking process begins with the majority voting strategy, followed by the successive integration of regression tests and reproduction tests. In addition, we conduct a comparative experiment using greedy sampling only to assess the effectiveness of the majority voting strategy.

\begin{table}[t]
\centering
\caption{Results of Component Ablation Study.}
\begin{tabular}{cccc}
\toprule
\textbf{System} & \textbf{\% Resolved} & \textbf{File Acc.} & \multicolumn{1}{l}{\textbf{Function Acc.}} \\ \midrule
\toolname{}    & 154 (51.3\%) & 81.2 & 52.4 \\
w/o Suggest & 114 (38.0\%) & 80.3 & 51.9 \\
w/o KG      & 132 (44.0\%) & 78.1 & 48.4 \\ \bottomrule
\end{tabular}
\label{table:ablation}
\end{table}

\begin{table}[t]
\centering
\caption{Results of Different Re-ranking Configurations.}
\begin{tabular}{cc}
\toprule
\textbf{Ranking Strategy} & \textbf{Claude-3.5-Sonnet} \\ \midrule
Greedy Sampling           & 113 (37.7\%)              \\
Majority Voting           & 121 (40.3\%)              \\
+ Regression Tests        & 134 (44.7\%)              \\
+ Reproduction Tests      & 154 (51.3\%)              \\ \bottomrule
\end{tabular}
\label{table:rerank}
\end{table}

\subsubsection{Results} 
Table \ref{table:ablation} summarizes the evaluation results of the full \toolname{} and its two ablated variants.
The variant w/o Suggest removes the suggestion module, while w/o KG removes the knowledge graph retrieval module.
Across all configurations, \toolname{} maintains a stable level of localization and repair performance, demonstrating that the overall framework inherently possesses a strong software repair capability.
However, removing individual components leads to varying degrees of performance degradation.

First, removing the Knowledge Graph module results in a slight drop in performance: the resolve rate decreases by 5.3\%, file-level localization accuracy by 3.1\%, and function-level localization accuracy by 4.0\%.
This suggests that without knowledge-graph-based retrieval, the agent struggles to fully comprehend the semantics of large-scale repository projects, which in turn weakens the performance of both the suggest and fix stages, ultimately reducing repair effectiveness.

In contrast, removing the Suggest module has a more pronounced impact on repair accuracy.
Although file-level and function-level localization accuracies decrease only marginally (0.9\% and 0.5\%), the overall resolve rate drops sharply by 13.3\%.
This confirms our earlier motivation: introducing a suggest stage significantly enhances semantic-level reasoning and cross-stage coordination.
The Suggest module effectively refines the agent’s exploration objective, improving repair success without compromising localization accuracy, and plays a critical role in \toolname{}’s superior performance on complex, repository-level repair tasks.

Table \ref{table:rerank} presents the ablation results of the Re-ranking module. As shown in the table, using only the majority voting strategy yields a resolve rate of 40.33\%, which already surpasses the greedy sampling baseline (37.67\%). When incorporating regression tests and reproduction tests into the re-ranking process, the resolve rate further increases to 44.67\% and 51.33\%, respectively.
These results highlight two key insights: (1) the majority voting strategy provides a more robust mechanism for selecting high-quality patches compared to simple greedy decoding, and (2) integrating regression and reproduction tests into the ranking process plays a crucial role in identifying the most correct and semantically consistent patch candidates.

\summary{2}{
\textbf{Findings:} (1) Removing the Knowledge Graph module leads to a 5.3\% decrease in resolve rate, and 3.1\% and 4.0\% reductions in file-level and function-level localization accuracy, respectively. (2) Removing the Suggest module results in only marginal declines in localization accuracy (0.9\% and 0.5\%), but causes a significant 13.3\% drop in resolve rate. (3) Within the Re-ranking module, the majority voting, regression tests, and reproduction tests contribute 6.6\%, 4.4\% and 2.6\% improvements to overall performance, respectively.  
\\ 
\textbf{Insights:} (1) All components play a meaningful role in the effectiveness of \toolname{}, with the Suggest module providing the most substantial contribution. (2) By introducing an additional optimization stage for refining the agent’s exploration direction, the suggest stage empowers \toolname{} to handle multi-hop reasoning and semantically entangled bugs more effectively, underscoring its crucial role in repository-level software repair.
}

\subsection{RQ3: Performance of Different Base Models}
\subsubsection{Design}
RQ3 investigates the performance of \toolname{} under different base LLMs.
We evaluate \toolname{} with four representative models: Claude-3.5-Sonnet~\cite{claude} (denoted as Claude-3.5), Claude-4-Sonnet~\cite{claude} (Claude-4), DeepSeek-V3~\cite{deepseekv3}, and Qwen3-235B-A22B~\cite{yang2025qwen3} (Qwen3).
For all experiments, we keep the configurations identical across models to ensure a fair comparison.
This experiment aims to evaluate the generalization capability of \toolname{} and examine how its performance varies with different base LLMs.

\begin{table}[t]
    \centering
    \caption{A comparison of \toolname{}'s performance on SWE-Bench-Lite using different base models.}
    \begin{tabular}{cccc}
        \toprule
        \textbf{System} & \textbf{\% Resolved} & \textbf{File Acc.} & \textbf{Function Acc.} \\
        \midrule
        Qwen3-235B-A22B   & 98 (32.7\%)  & 71.0 & 45.3 \\
        DeepSeek-V3       & 113 (37.7\%) & 77.7 & 51.7 \\
        Claude-3.5-Sonnet & 154 (51.3\%) & 81.2 & 52.4 \\
        Claude-4-Sonnet   & 182 (60.7\%) & 88.7 & 57.2 \\
        \bottomrule
    \end{tabular}
    \label{table:dif_llm}
\end{table}

\subsubsection{Results}
Table~\ref{table:dif_llm} reports the performance of \toolname{} with four different base LLMs.
Overall, stronger base models consistently lead to better repair and localization performance.
Among the evaluated models, Claude-4 achieves the best overall results, followed by Claude-3.5, DeepSeek-V3, and Qwen3.

Specifically, Claude-4 resolves 182 instances, achieving a 60.7\% resolution rate, with file-level and function-level localization accuracies of 88.7\% and 57.2\%, respectively.
Under the same base model, this result also surpasses the Claude-4-based baselines reported in RQ1, including KGCompass (58.3\%) and ExpeRepair-v1.0 (60.3\%), validating the superiority of our locate-suggest-fix architecture in leveraging highly capable LLMs.
Claude-3.5 ranks second, resolving 154 instances (51.3\%), with 81.2\% file-level and 52.4\% function-level localization accuracy.
DeepSeek-V3 achieves a 37.7\% resolution rate, together with 77.7\% file-level and 51.7\% function-level accuracy, while Qwen3 performs the weakest, with 32.7\%, 71.0\%, and 45.3\% on the three metrics, respectively.
Compared with the gaps in resolution rate, the differences across models are smaller on localization accuracy.
This suggests that the locate stage is relatively less sensitive to the capability of the base LLM, whereas the suggest and fix stages depend more heavily on the model’s reasoning, synthesis, and instruction-following abilities.
In particular, the suggest stage requires the model to integrate retrieved context with the issue description, refine the repair objective, and generate feasible repair strategies, which places higher demands on semantic reasoning than localization alone.
Overall, these results suggest that \toolname{} generalizes well across different base LLMs, while its final performance still benefits substantially from stronger underlying models.

\summary{3}{
\textbf{Findings:} 
(1) Claude-4-Sonnet achieves the best overall performance, with a 60.7\% resolve rate, 88.7\% file-level accuracy, and 57.2\% function-level accuracy.
(2) Claude-3.5-Sonnet ranks second, attaining 51.3\%, 81.2\%, and 52.4\% on the three metrics, respectively.
(3) DeepSeek-V3 and Qwen3-235B-A22B achieve lower performance, with Qwen3 showing the weakest overall results. 
\\
\textbf{Insights:} (1) \toolname{} exhibits consistent and robust performance across different base LLMs, highlighting its strong generalization ability and methodological stability. (2) Among all system stages, the suggest stage shows the highest dependency on model capability. (3) \toolname{} is largely model-agnostic, maintaining its effectiveness and reliability across diverse language model architectures.
}

\subsection{RQ4: Generalization Capability in Vulnerability Repair}
\subsubsection{Design} 

To verify the generality of \toolname{}, we extend our framework to the vulnerability repair domain and develop \toolname{}-Vul.
We evaluate \toolname{}-Vul on two widely used datasets, including VUL4J~\cite{vul4j} and VJBench~\cite{effective}. 
We include all 15 vulnerabilities from VJBench in our evaluation.
Given the test-oriented nature of these benchmarks, we further incorporate a failure-guided patch enhancement mechanism:
When a generated patch fails to pass validation, the failure test information is fed back as additional contextual input to the fixer for re-repair. 
If the fixer fails three consecutive times, the failed context is passed to the suggester for re-suggestion; after two failed iterations, it is escalated to the localizer for re-localization.
To prevent infinite repair loops, we set an upper limit of 100 repair iterations per instance.

We compare \toolname{}-Vul against three state-of-the-art vulnerability repair methods, FSV~\cite{effective}, NTR~\cite{ntr}, VRPILOT\cite{vrpilot}, and two advanced general-purpose software engineering agents SWE-Agent\cite{sweAgent} and OpenHands\cite{openhandsArticle}, using vulnerability resolve rate as the primary evaluation metric.
FSV is the first work to study and compare Java vulnerability repair capabilities of LLMs and DL-based APR models. NTR combines the strengths of both templates and large-scale LLMs to fix Java vulnerabilities. VRPILOT uses a chain-of-thought prompt to reason about a vulnerability prior to generating patch candidates and iteratively refines prompts according to the output of external tools on previously-generated patches.
SWE-Agent is a general software engineering agent designed for repository-level problem solving through a structured command-line interaction interface. OpenHands is a general agent framework that supports tool use, terminal interaction, and multi-step reasoning in software engineering tasks.

Additionally, we conduct ablation studies on \toolname{}-Vul by removing the Suggest module and Knowledge Graph module, respectively, and by substituting the base model with GPT-4o, DeepSeek-V3, and Qwen3-235B-A22B.
Notably, since the failure-guided enhancement mechanism introduces additional computational cost, we adopt GPT-4o as a more cost-efficient alternative for the base model, compared to Claude-3.5-Sonnet.

\begin{table*}[t]
    \centering
    \footnotesize
    \caption{Comparison results between \toolname{}-Vul and existing baselines on VUL4J and VJBench.}
    \begin{tabular}{c|ccccccc}
        \toprule
        \textbf{} & \textbf{\toolname{}-Vul} & \textbf{FSV-Codex} & \textbf{FSV-finetuned} & \textbf{NTR} & \textbf{VRPILOT} & \textbf{SWE-Agent} & \textbf{OpenHands} \\
        \midrule
        \textbf{VUL4J}   & 17 (48.6\%) & 11 (31.1\%) & 9 (25.7\%)  & 14 (40.0\%) & 14 (40.0\%) & 11 (31.4\%) & 14 (40.0\%) \\
        \textbf{VJBench} & 7 (46.7\%)  & 6 (40.0\%)    & 4 (26.7\%)  & N/A         & 6 (40.0\%)  & N/A                   & N/A \\
        \bottomrule
    \end{tabular}
    \label{table:vul}
\end{table*}

\begin{table*}[t]
    \centering
    \footnotesize
    \caption{Comparison of vulnerability repair rates and localization accuracies across different \toolname{}-Vul variants.}
    \begin{tabular}{c|ccccc}
        \toprule
        \textbf{Configuration} & \textbf{\toolname{}-Vul} & \textbf{w/o Suggest} & \textbf{w/o KG} & \textbf{DeepSeek-V3} & \textbf{Qwen3-235B-A22B} \\
        \midrule
        \textbf{\% Resolved}  & 24 (48.0\%) & 13 (26.0\%) & 11 (22.0\%) & 26 (52.0\%) & 20 (40.0\%) \\
        \textbf{File Acc.}    & 72.0        & 68.0        & 66.0        & 76.0        & 76.0 \\
        \textbf{Function Acc.}& 64.0        & 44.0        & 42.0        & 62.0        & 54.0 \\
        \bottomrule
    \end{tabular}
    \label{table:vul_dif}
\end{table*}

\subsubsection{Results}
Table \ref{table:vul} shows the comparison results between \toolname{}-Vul and the baselines on VUL4J and VJBench. We report the number of applied patches on 35 vulnerabilities from VUL4J and 15 from VJBench. $FSV_{Codex}$ denotes FSV with zero-shot Codex model and $FSV_{Incoder\_finetuned}$ denotes FSV fine-tuned with general APR data. 
Overall, \toolname{}-Vul achieves the best performance among the compared methods on both benchmarks, resolving 17 vulnerabilities on VUL4J and 7 on VJBench. On VUL4J, \toolname{}-Vul outperforms all specialized vulnerability repair baselines, including FSV-finetuned (9), NTR (14), and VRPILOT (14). 
It also performs better than the two recent general-purpose software engineering agents, SWE-Agent (11) and OpenHands (14). On VJBench, \toolname{}-Vul again achieves the best result, surpassing FSV-Codex (6), FSV-finetuned (4), and VRPILOT (6).
These comparisons suggest that \toolname{}-Vul is effective not only against prior vulnerability-repair methods, but also against recent general-purpose agent frameworks.

Table \ref{table:vul_dif} presents the comparison of vulnerability resolution rates and localization accuracies across different \toolname{}-Vul variants.
Here, \toolname{}-Vul denotes the full approach using GPT-4o as the base model; w/o Suggest and w/o KG represent the variants with the Suggest and Knowledge Graph modules removed, respectively; DeepSeek-V3 and Qwen3-235B-A22B denote variants using corresponding base models.

To evaluate the effectiveness of each component, we remove either the Suggest or Knowledge Graph module, which leads to a substantial drop in performance.
The repair rate decreases from 48.0\% to 26.0\% and 22.0\%, respectively.
File-level and function-level localization accuracies drop from 72.0\% and 68.0\% to 68.0\% / 44.0\% (without Suggest) and 66.0\% / 42.0\% (without Knowledge Graph).
These results are consistent with the findings in RQ2, further confirming the critical roles of the Suggest and Knowledge Graph modules in achieving effective repair.
Moreover, a cross-comparison between RQ2 and RQ4 reveals that the Knowledge Graph module plays a relatively greater role in vulnerability repair, whereas the Suggest module contributes more significantly in software repair.
Additionally, the Suggest module also impacts localization accuracy due to the failure-guided patch enhancement mechanism, which aligns with our design expectations.

Regarding base model performance, DeepSeek-V3 achieves the best overall results, with a 52.0\% repair rate, 76.0\% file-level accuracy, and 62.0\% function-level accuracy, outperforming GPT-4o (48.0\%, 72.0\%, 64.0\%) and Qwen3-235B-A22B (40.0\%, 76.0\%, 54.0\%).
This performance trend is consistent with that observed in RQ3, further demonstrating the strong generalization ability of the \toolname{} framework across different domains and base models.

\summary{4}{
\textbf{Findings:} (1) \toolname{}-Vul outperforms all baseline approaches in vulnerability repair scenarios, achieving a 48.0\% resolve rate, 72.0\% file-level, and 64.0\% function-level localization accuracy.
(2) Removing the Suggest or Knowledge Graph module leads to substantial performance degradation - the resolve rate drops by 22.0\% and 26.0\%, file-level localization accuracy decreases by 4.0\% and 6.0\%, and function-level localization accuracy falls by 20.0\% and 22.0\%, respectively.
(3) Among different base LLMs, DeepSeek-V3 achieves the best results (52.0\%, 76.0\%, and 62.0\% for the three metrics), followed by GPT-4o (48.0\%, 72.0\%, and 64.0\%) and Qwen3-235B-A22B (40.0\%, 76.0\%, and 54.0\%). 
\\
\textbf{Insights:} (1) When extended to the vulnerability repair domain, \toolname{} maintains strong repair performance, demonstrating robust generalization ability. (2) The consistency of results between software repair and vulnerability repair tasks further validates \toolname{}’s robustness, practical applicability, component effectiveness, and model-agnostic design, confirming its potential for broader real-world adoption.
}

\section{Discussion}
\subsection{Comparison on SWE-Bench-Verified}
To provide a more rigorous evaluation and reduce potential bias caused by unverified or ill-formed issues, we further evaluate \toolname{} on the SWE-Bench-Verified dataset.
SWE-Bench-Verified is a strictly human-validated subset of 500 instances, in which each issue is confirmed to be well-posed and unambiguously solvable.

Table~\ref{tab:rq1_extra} presents the comparison between \toolname{} and recent state-of-the-art approaches on SWE-Bench-Verified, all using Claude-3.5-Sonnet as the base model. The results show that \toolname{} achieves the best performance among the compared approaches on this rigorous benchmark. Specifically, \toolname{} successfully resolves 327 out of 500 instances, achieving a resolution rate of 65.4\%. This clearly outperforms several strong baselines, including EPAM AI (62.8\%), AutoCodeRover-v2.1 (51.6\%), Agentless-1.5 (50.8\%), and SWE-Agent (33.6\%).

\begin{table*}[t]
  \centering
  \footnotesize
  \caption{A comparison of repair results between \toolname{} and existing baselines on SWE-Bench-Verified}
  \resizebox{0.7\textwidth}{!}{
    \begin{tabular}{ccc}
    \toprule
    \textbf{Approaches} & \textbf{Base Model } & \textbf{\% Resolved} \\
    \midrule
    SWE-Agent & Claude-3.5-Sonnet & 168(33.6\%) \\
    Agentless-1.5 & Claude-3.5-Sonnet & 254(50.8\%) \\
    AutoCodeRover-v2.1 & Claude-3.5-Sonnet & 258(51.6\%) \\
    OpenHands + CodeAct v2.1 & Claude-3.5-Sonnet & 265(53.00\%) \\
    \toolname{} & Claude-3.5-Sonnet & 327(65.4\%) \\
    \bottomrule
    \end{tabular}%
    }
  \label{tab:rq1_extra}%
\end{table*}%

This substantial performance margin on the Verified dataset highlights a critical advantage of our locate-suggest-fix paradigm. While procedural or purely trial-and-error baselines (such as Agentless or SWE-Agent) may struggle with complex cross-file dependencies even when the issue itself is well-posed, \toolname{} leverages the Suggest stage and the Knowledge Graph toolkit to maintain deep reasoning consistency. By generating explicit, coordinated repair strategies before code modification, \toolname{} effectively capitalizes on the high-quality nature of the Verified instances, translating clear problem descriptions into highly accurate, repository-level patches.

\subsection{Case study: The Role of the Suggester}
\begin{figure*}[t]
    \centering
    \includegraphics[width=1.0\linewidth]{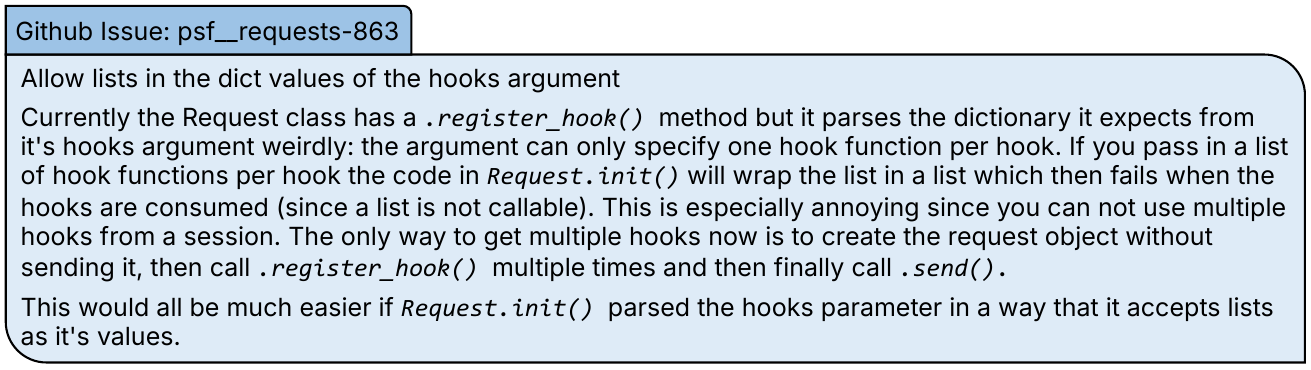}
    \caption{The issue \texttt{psf\_\_requests-863} from SWE-Bench-lite}
    \Description{Example of an issue from SWE-Bench-lite.}
    \label{fig:issue}
\end{figure*}

To explicitly illustrate how suggester bridges the gap between localization and repair, we present a qualitative case study on the issue \texttt{psf\_\_requests-863} from SWE-Bench.
As shown in Figure~\ref{fig:issue}, the core problem is that the Request object fails to properly handle the list of hook functions passed during its initialization.

\begin{figure*}[t]
    \centering
    \includegraphics[width=1\linewidth]{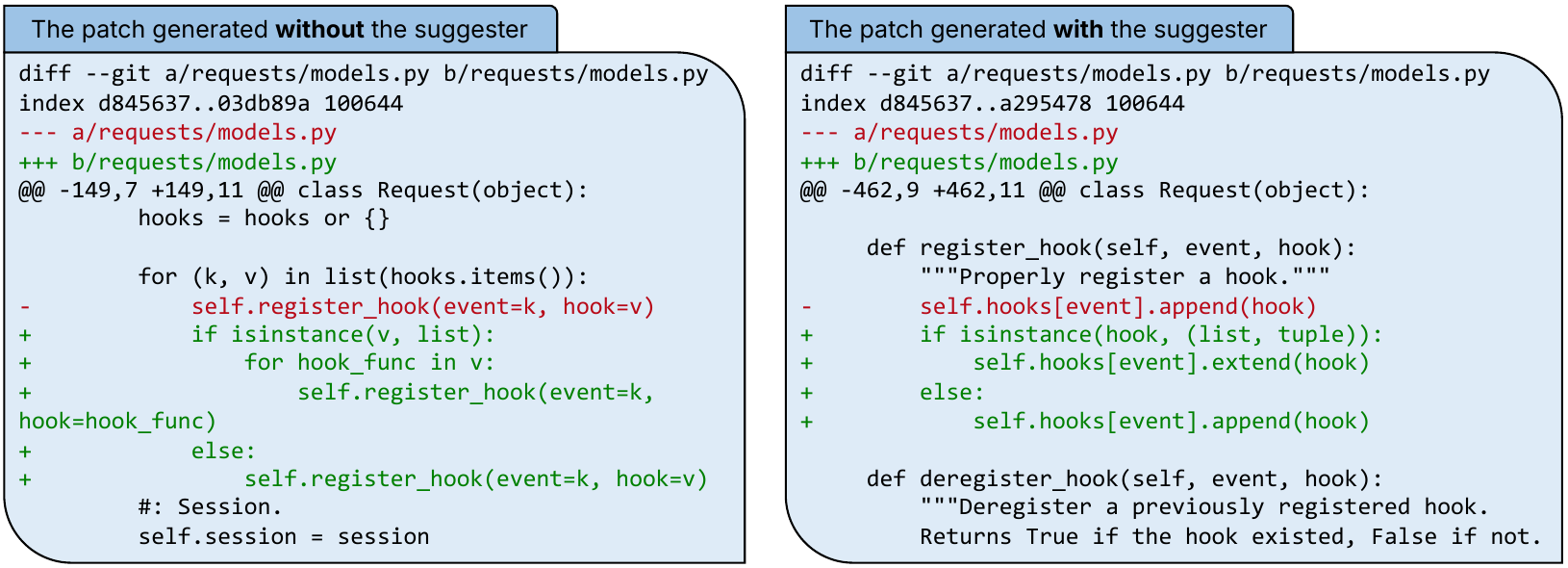}
    \caption{Comparison of psf\_\_requests-863 patches with/without the suggester agent. Left: the patch generated without the suggester; right: the patch generated with the suggester.}
    \Description{Comparison of patches: left shows patch without suggester agent, right shows patch with suggester agent.}
\label{fig:compare}
\end{figure*}

\begin{figure*}[t]
    \centering
    \includegraphics[width=0.8\linewidth]{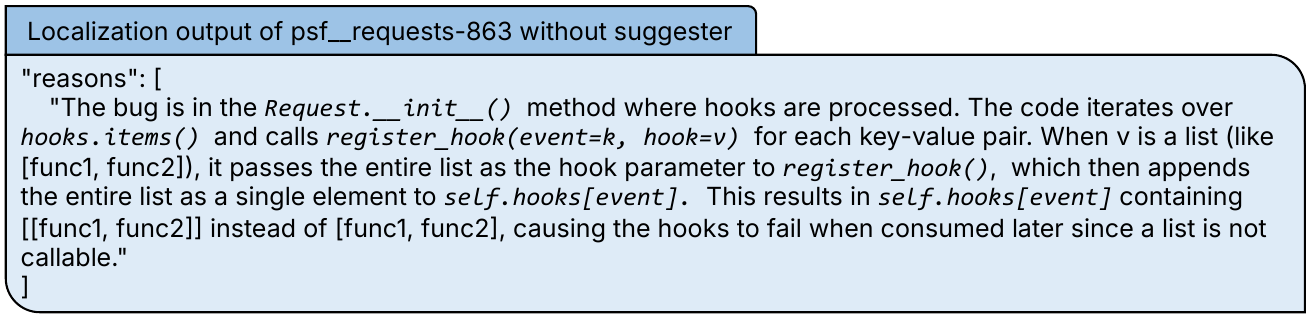}
    \caption{Localization output of psf\_\_requests-863 without suggester.}
    \Description{Localization output of psf\_\_requests-863 without suggester.}
    \label{fig:wo_suggest}
\end{figure*}

Figure~\ref{fig:compare} presents the patches generated by \toolname{} with and without the suggester agent.
In a standard localize-then-fix pipeline (\ie without the suggester), the agent is highly susceptible to the localized and often misleading hints in the raw issue description.
As is shown in Figure~\ref{fig:wo_suggest}, the original issue report explicitly points to the initialization phase and states that ``This would all be much easier if \texttt{Request.\_init\_()} parsed the \texttt{hooks} parameter in a way that it accepts lists as its values.''
Constrained by this initial clue, the baseline agent follows the user’s naive suggestion and patches only the apparent symptom by hardcoding a type check (\texttt{isinstance(v, list)}) inside the initialization loop of \texttt{Request.\_init\_()}. 
However, this superficial fix fails the repository’s validation tests. Because it limits the change to the initialization phase and ignores standalone \texttt{register\_hook} calls, it is functionally incomplete. 
Moreover, it violates the single-responsibility principle and supports only lists, not tuples, demonstrating that localized trial-and-error based on imperfect issue descriptions can produce plausible yet incorrect patches.

\begin{figure*}[t]
    \centering
    \includegraphics[width=0.8\linewidth]{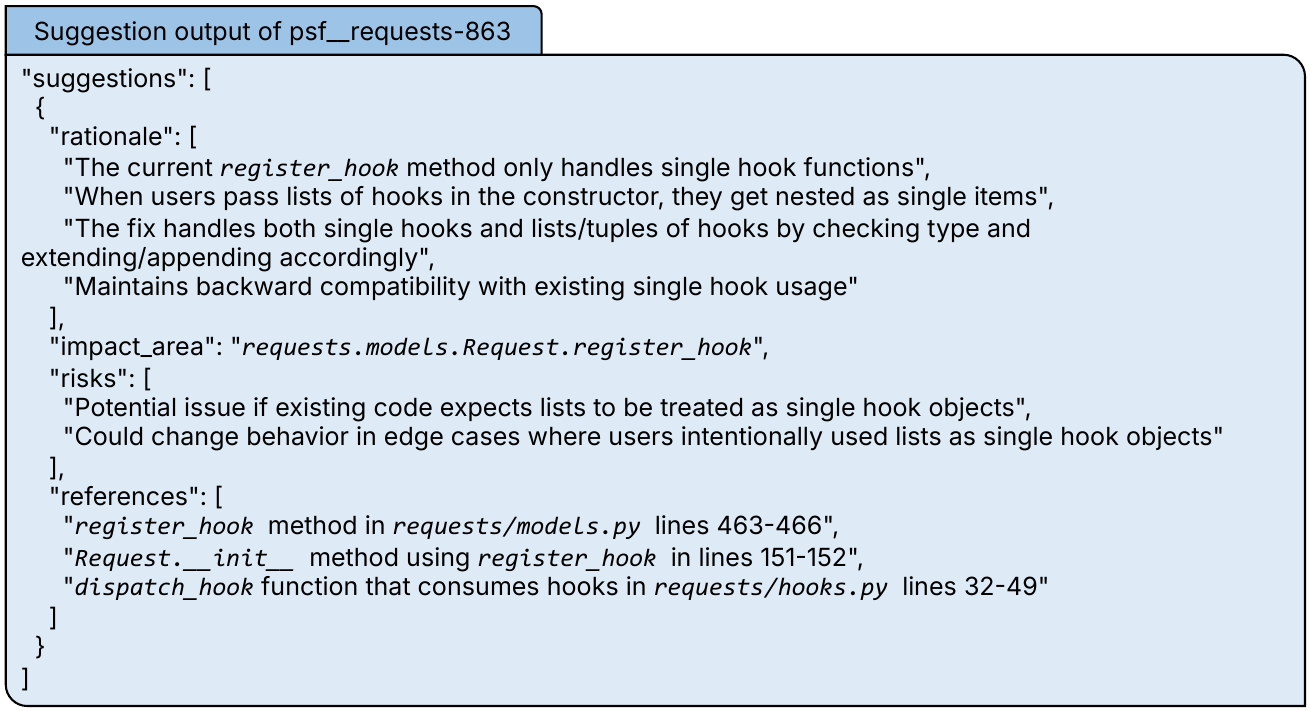}
    \caption{Suggestion output of psf\_\_requests-863.}
    \Description{Suggestion output of psf\_\_requests-863.}
    \label{fig:w_suggest}
\end{figure*}

In contrast, the introduction of the suggester fundamentally shifts the paradigm from local trial-and-error to global architectural planning. By leveraging the knowledge graph for backward analysis, the suggester does not blindly patch \_\_init\_\_(). Instead, its generated output explicitly redefines the true impact area to the underlying register\_hook method. As is shown in Figure~\ref{fig:w_suggest}, A typical suggester output provides the fixer with a clear rationale and an actionable blueprint: "The current register\_hook method only handles single hook functions.
The fix should handle both single hooks and lists/tuples by checking the type and extending/appending accordingly to maintain backward compatibility."

Guided by this explicit and structured plan, the fixer completely abandons the flawed \_\_init\_\_ modifications. It directly targets the register\_hook function, generating a highly cohesive patch. Unlike the baseline, the suggester-guided patch successfully passes all validation tests. It supports both lists and tuples, perfectly covers all hook registration scenarios, maintains strict backward compatibility, and adheres to expert-level design principles. This qualitative comparison clearly demonstrates that the suggester is an indispensable planning module, successfully preventing the LLM from falling into the trap of localized, fragmented edits that ultimately fail execution.

\subsection{Tool Usage Analysis}

To provide further insight into \toolname{}'s behavior and its interactions with the knowledge graph, we record the frequency of all tool invocations during the SWE-Bench-Lite evaluation.
As shown in Table~\ref{tab:tool}, \toolname{} makes use of the entire KG-based toolkit.
All 14 designed tools are invoked at least once, suggesting that the toolkit supports a broad range of repair scenarios and that even specialized tools are useful in specific cases.

\begin{table}[t]
    \centering
    \footnotesize
    \caption{Invoke frequency of \toolname{}'s toolkit.}
    \begin{tabular}{lcc}
        \toprule
        \textbf{Tools} & \textbf{Invoked Times} & \textbf{Percentage} \\
        \midrule
        analyze\_file\_structure                     & 289  & 3.25\%  \\
        get\_code\_relationships                    & 25   & 0.28\%  \\
        find\_methods\_by\_name                     & 399  & 4.49\%  \\
        extract\_complete\_method                   & 421  & 4.74\%  \\
        find\_class\_constructor                    & 98   & 1.10\%  \\
        list\_class\_attributes                     & 13    & 0.14\%  \\
        find\_variable\_usage                       & 54   & 0.61\%  \\
        find\_all\_variables\_named                 & 14   & 0.16\%  \\
        show\_file\_imports                         & 97   & 1.10\%  \\
        search\_code\_with\_context                 & 1243 & 13.99\% \\
        find\_files\_containing                     & 733  & 8.24\%  \\
        explore\_directory                          & 538  & 6.05\%  \\
        read\_file\_lines                           & 4512 & 50.78\% \\
        execute\_shell\_command\_with\_validation   & 450  & 5.06\%  \\
                \midrule
        Total & 8886 & 100\% \\
        \bottomrule
    \end{tabular}
    \label{tab:tool}
\end{table}

In total, \toolname{} makes 8,886 tool calls, and the resulting usage pattern follows a clear long-tail distribution.
Among all tools, \textit{read\_file\_lines} is used most frequently, with 4,512 invocations.
This result is consistent with the architectural design of our multi-agent framework: rather than relying solely on the LLM's internal memory, the agents repeatedly query the repository's exact physical state to verify line numbers and local code context before generating patches.
Following \textit{read\_file\_lines}, the most frequently used tools are those supporting broader semantic exploration, including \textit{search\_code\_with\_context} (1,243 calls), \textit{find\_files\_containing} (733 calls), and \textit{explore\_directory} (538 calls).
These tools are heavily used by the localizer and suggester to explore repository context, trace cross-file dependencies, and support the backward analysis needed for coordinated repair planning.
In addition, the substantial use of \textit{execute\_shell\_command\_with\_validation} (450 calls) reflects the framework's interaction with the execution environment, allowing the agents to inspect repository states and validate hypotheses when necessary.
By contrast, fine-grained entity-analysis tools (\eg, \textit{get\_code\_relationships} and \textit{list\_class\_attributes}) appear in the long tail of the distribution and are typically invoked only when the agents encounter highly specific object-oriented structural constraints.
Overall, the observed distribution suggests that, although the individual tools in \toolname{} are lightweight, all pre-defined APIs are effectively utilized in practice and together provide sufficient support for software repair across the localization, suggestion, and fixing stages.

\section{Related Work}
\subsection{LLM-based Automated Program Repair}
Recent advances in LLMs have fundamentally reshaped the landscape of APR~\cite{zhang2024systematic, hu2025can}. Although earlier learning-based approaches have demonstrated promising results~\cite{zhang2023surveyapr, jiang2021cure, zhang2023gamma, li2022improving, xia2023automated}, their reliance on large-scale labeled data and limited generalization across languages and repositories constrain their applicability in real-world settings. In contrast, LLM-based methods leverage the inherent reasoning and code understanding abilities of foundation models, enabling repair without task-specific retraining. This paradigm shift has led to a surge of research exploring diverse prompting and reasoning strategies, including zero-shot~\cite{zeroshot1,zeroshot2,zeroshot3, zero_shot}, few-shot~\cite{fewshot1, fewshot2, fewshot3} prompting, to elicit repair behaviors directly from pretrained models. 

Retrieval-Augmented Generation (RAG) techniques~\cite{ragfix, knowledge} further enhance these models by grounding their reasoning in external knowledge sources, improving contextual precision in complex repositories. Building on these advances, recent works have begun to move beyond static prompting toward interactive, agent-driven repair~\cite{sweAgent, autocoderover, design}, where LLMs dynamically interact with tools, files, and execution environments to iteratively refine their reasoning and patch generation. Such approaches mark a transition from single-turn prediction to multi-step decision-making, bridging code understanding, localization, and fix generation in a unified, autonomous framework.

Beyond prompt-based methods, fine-tuning and Reinforcement Learning (RL) have emerged as complementary strategies for enhancing LLMs’ software repair capability. Fine-tuning approaches (\eg DeepDebug~\cite{deepdebug}, CIRCLE~\cite{circle}, APRFiT~\cite{aprfix}) typically adapt a pre-trained LLM to domain-specific repair corpora, allowing the model to internalize recurring bug–fix patterns and language-specific conventions. However, such methods often suffer from high computational cost, catastrophic forgetting, and limited adaptability to unseen projects or languages. In contrast, RL offers a dynamic optimization paradigm that directly aligns model behavior with repair success signals. By treating code repair as a sequential decision-making process, RL-based approaches, such as SWE-RL~\cite{swerl}, RepairLLaMA~\cite{repairllama} and Repair-R1~\cite{repairr1}, allow models to iteratively refine their outputs based on execution feedback or test results. These methods enhance the model’s ability to reason about repair quality beyond static supervision.

In this work, we leverage existing LLMs to address complex repository-level software repair tasks. To minimize training cost and methodological complexity, we adopt a train-free multi-agent framework, enabling \toolname{} to autonomously locate and repair bugs. 

\subsection{Repository-level Software Repair}

With the rapid growth of modern software projects, repository-level software repair has become an increasingly challenging task. Pure LLM-based approaches often struggle with multi-step reasoning and complex context management~\cite{evaluatingCodex, evaluatingSynthesis, evaluatingTools}, motivating the exploration of new paradigms for large-scale program repair. Recent advances have popularized the agentic paradigm for repository-level repair~\cite{liu2023dynamic, zhao2024expel, talebirad2023multi, schmidgall2025agent, wang2024executable}. 
Agent-based systems such as AutoGPT~\cite{AutoGPT}, GPT Engineer~\cite{gpt-engineer} and OpenHands~\cite{openhandsArticle} demonstrate that tool-augmented agents can iteratively analyze, plan, and act to solve complex software repair tasks. These agents perform repository reasoning, generate candidate patches, and validate fixes through automated testing\cite{codemodelimpact, codeagent, vulrepair, chatgpt4vul, graphcoder}.
For example,
RepairAgent~\cite{repairagent} introduces the first work to address the software repair challenge through an autonomous agent based on LLMs.
SWE-Agent~\cite{sweAgent} integrates multi-step retrieval and testing to repair repository-level software bugs.
OpenHands~\cite{openhandsArticle} introduces a multi-agent collaboration design for software repair and patch validation.
AutoCodeRover~\cite{autocoderover} works on a program representation (abstract syntax tree) as opposed to viewing a software project as a mere collection of files.
VRPILOT~\cite{vrpilot} uses a chain-of-thought prompt to reason about a vulnerability prior to generating patch candidates and iteratively refines prompts according to the output of external tools on previously-generated patches.
In contrast, several procedure-based approaches rely on carefully designed static pipelines to organize and guide LLMs during repository-level repair.
For example, Agentless~\cite{agentless} employs a simple three-phase process, \ie localization, repair, and validation, without allowing the LLM to autonomously plan or interact with complex toolchains. 
Similarly, Moatless~\cite{antoniades2024swe} argues that strong tool integration and sufficient contextual grounding can outperform fully autonomous agentic control.

Unlike prior work that primarily treats repair as a localize-then-fix process, \toolname{} introduces a suggestion-guided multi-agent framework that explicitly separates bug repair into three stages: locate, suggest, and fix.
Although some existing approaches incorporate reasoning mechanisms such as chain-of-thought reasoning, such reasoning is typically tightly coupled with patch generation and remains largely unguided, leaving the model to determine the modification strategy on its own. 
In contrast, \toolname{} is designed to mimic real-world developer practice by explicitly decomposing repository-level repair into specialized roles and enabling collaboration among multiple agents. 
In particular, the key novelty of \toolname{} lies in introducing the Suggester as an explicit intermediate stage between localization and repair. 
Rather than directly editing localized snippets, the Suggester first performs secondary context exploration and dependency analysis based on candidate bug locations, and then transforms coarse localization signals into actionable, repository-aware repair suggestions before any code is modified. 
Together with the knowledge graph and specialized toolkit, this design allows \toolname{} to bridge the reasoning gap between ``where the bug is'' and ``how the code should be changed.''

\subsection{Knowledge-Graph-Guided Repository Retrieval}
Knowledge graphs have recently become a powerful tool for enhancing large-scale code understanding and repository-level reasoning~\cite{repograph, KGCompass, graphcoder, liu2024codexgraph,chen2025prometheus}. By representing software repositories as structured graphs of entities (\eg classes, methods, and variables) and their relations (\eg inheritance, invocation, and data dependencies), knowledge-graph-based approaches enable more precise and semantically grounded retrieval of relevant contexts for program repair and comprehension tasks.
RepoGraph~\cite{repograph} proposes a plug-in module that manages a repository-level structure for modern AI software engineering solutions. KGCompass~\cite{KGCompass} proposes a novel repository-aware knowledge graph that accurately links repository artifacts (issues and pull requests) and codebase entities (files, classes, and functions) and a path-guided repair mechanism that leverages KG-mined entity paths, tracing through which allows us to augment LLMs with relevant contextual information to generate precise patches along with their explanations. 
GraphCoder~\cite{graphcoder} leverages control-flow, data- and control-dependence between code statements to build a retrieval-augmented code completion framework. 
CodexGraph~\cite{liu2024codexgraph} enables the LLM agent to construct and execute queries, allowing for precise, code structure-aware context retrieval and code navigation by leveraging the structural properties of graph databases and the flexibility of the graph query language.

Unlike prior work that mainly use knowledge graphs as static contextual augmentation, \toolname{} operationalizes repository knowledge through a scenario-oriented toolkit for dynamic repository-level reasoning. 
This distinction is particularly important when comparing \toolname{} with recent graph-based methods such as KGCompass~\cite{KGCompass}. 
KGCompass adopts a path-guided mechanism that relies on pre-computed entity paths mined from the knowledge graph and injects them into prompts as supplementary context. 
In contrast, \toolname{} encapsulates repository knowledge into a scenario-oriented API toolkit comprising 14 specialized tools. 
Rather than passively consuming pre-computed paths, the agents in \toolname{} invoke different tools according to the task context at runtime. 
As a result, \toolname{} supports dynamic, fine-grained, and scenario-specific interaction with repository knowledge, making retrieval more precise, flexible, and adaptive for repair-oriented reasoning.

\section{Threats to Validity}
\subsection{Internal Validity}
Internal validity concerns potential experimental biases that may affect the fairness or consistency of our evaluation. Since LLMs inherently involve stochastic generation, randomness in model outputs may introduce noise into the reported results. To mitigate this threat, we employ a multi-sampling strategy: for each instance, the model generates four candidate localization sets, and for each localization, patches are sampled once with temperature = 0 and four times with temperature = 0.8. This design balances stability and exploration, reducing the variance caused by random sampling while ensuring fair coverage of \toolname{}’s repair potential. Furthermore, all experiments are conducted under identical configurations, with fixed random seeds and isolated environments, minimizing uncontrolled sources of variation.

\subsection{External Validity}
External validity addresses the generalizability of our findings beyond the specific datasets and programming languages used in this study. Our primary experiments are conducted on Python-based projects from the SWE-Bench-Lite benchmark, which may raise concerns about language dependency. To address this, we extend our evaluation to Java-based vulnerability datasets (\eg, VUL4J, VJBench), and observe consistent repair rates and localization accuracies across both settings. These results demonstrate that \toolname{}’s architecture and methodology exhibit strong cross-language generalizability, indicating that its mechanisms for locate, suggest and fix stages are not restricted to a particular programming language or dataset domain.

\subsection{Construct Validity}
Construct validity concerns whether our evaluation metrics accurately capture what they are intended to measure, \ie the bug localization and repair capability of \toolname{}, rather than artifacts of the underlying language model. 
\toolname{} is primarily implemented using Claude-3.5-Sonnet as the base model, which may introduce biases tied to model-specific performance. 
To mitigate this, we further evaluate \toolname{} using DeepSeek-V3 and Qwen3-235B-A22B. The consistent performance trends observed across these LLMs demonstrate that \toolname{}’s improvements stem from its framework design rather than from the characteristics of any specific model. This supports the validity of our claim that \toolname{}’s effectiveness is model-agnostic and method-driven.

\section{Conclusion}
\label{sec:conclusion}
In this paper, we introduce \toolname{}, a suggestion-guided multi-agent framework designed to automatically analyze and repair repository-level software bugs. \toolname{} consists of three cooperating agents: localizer, suggester, and fixer, which work together to understand the structure of complex repositories and propose effective patches. \toolname{} introduces a novel suggestion stage into the existing automated software repair pipeline.
By performing secondary contextual refinement based on the exploratory results of the localization stage, \toolname{} demonstrates strong effectiveness when handling complex repositories.
Furthermore, we equip \toolname{} with a repository-level knowledge graph and a corresponding retrieval toolkit, which together enhance LLMs' ability to comprehend the structural and semantic relationships within large-scale codebases.
Our evaluation demonstrates that \toolname{} outperforms existing state-of-the-art approaches on resolve rate, file-level localization accuracy, and function-level localization accuracy while maintaining a relatively low cost. 
In addition, \toolname{} exhibits strong scalability across diverse LLM architectures and maintains robust performance across different programming languages, highlighting its practical applicability and potential for deployment in real-world software development environments.

\section*{Acknowledgments}

This work is supported partially by Natural Science Foundation of Jiangsu Province (BK20251458), National Natural Science Foundation of China (U24A20337, 62372228), and Frontier Technologies R\&D Program of Jiangsu (BF2024070).

\bibliographystyle{ACM-Reference-Format}
\bibliography{reference}

\end{document}